\documentclass[letterpaper]{article} 
\usepackage{aaai2026}  
\usepackage{times}  
\usepackage{helvet}  
\usepackage{courier}  
\usepackage[hyphens]{url}  
\usepackage{graphicx} 
\usepackage{natbib}  
\usepackage{caption} 
\usepackage{algorithmic}

\usepackage{times}
\usepackage{soul}
\usepackage{url}
\usepackage[utf8]{inputenc}
\usepackage{graphicx}
\usepackage{amsmath}
\usepackage{amssymb}
\usepackage{amsthm}
\usepackage{booktabs}
\usepackage{comment}
\usepackage{multirow}

\usepackage[T1]{fontenc}

\usepackage{stmaryrd}
\usepackage{mathtools}
\usepackage[geometry]{ifsym}
\usepackage{paralist}
\usepackage{makecell}
\usepackage{color, colortbl}
\usepackage{multirow}
\usepackage{multicol}
\usepackage{mathrsfs}
\usepackage{caption}
\usepackage{subcaption}

\usepackage{thm-restate}

\usepackage{centernot}

\usepackage{lineno}

\usepackage[dvipsnames]{xcolor}

\usepackage{cleveref}

\usepackage[ruled,vlined,linesnumbered]{algorithm2e}
\Crefname{algocf}{Algorithm}{Algorithms}

\usepackage{pgfplots}
\usepackage{tikz}
\pgfplotsset{width=10cm,compat=1.9}
\usetikzlibrary{arrows,backgrounds,automata,topaths,patterns,decorations.pathreplacing,decorations.pathmorphing}
\usetikzlibrary{arrows, chains, positioning, quotes, shapes.geometric}
\usetikzlibrary{calc}
\usetikzlibrary{fit}
\usetikzlibrary{arrows}
\usetikzlibrary{positioning,automata}

\newcommand{\ST}{ST}

\newcommand{\ALC}{\ensuremath{\mathcal{ALC}}}

\newcommand{\ALCOiu}{\ensuremath{\mathcal{ALCO}_u^\iota}}

\newcommand{\ALCi}{\ensuremath{\mathcal{ALC}\iota}}
\newcommand{\ALCig}{\ensuremath{\mathcal{ALC}\iota_G}}
\newcommand{\ALCil}{\ensuremath{\mathcal{ALC}\iota_L}}

\newcommand{\ALCii}{\ensuremath{\mathcal{ALC}\iota}}

\newcommand{\FOtwo}{\ensuremath{\textup{FO}^2}}

\newcommand{\T}{\mathcal{T}}

\newcommand{\I}{\mathcal{I}}
\newcommand{\J}{\mathcal{J}}
\newcommand{\Ont}{\mathcal{O}}

\newcommand{\Lan}{\mathcal{L}}

\newcommand{\Dom}{Dom}
\newcommand{\Rng}{Rng}

\newcommand{\names}{\textit{Names}}
\newcommand{\named}{\textit{NamedInd}}

\newcommand{\maxB}{\ensuremath{\textsc{MaxBsimALC}}}

\newcommand{\EXPT}{\ensuremath{\textsc{\textup{ExpTime}}}}

\usepackage{todonotes}

\definecolor{applegreen}{rgb}{0.55, 0.71, 0.0}
\definecolor{cadmiumred}{rgb}{0.89, 0.0, 0.13}

\newcommand{\theory}[2]{Th_{#1}(#2)}
\newcommand{\TALCi}{\ensuremath{\mathtt{TAB}_{\mathcal{ALC}\iota}}}
\newcommand{\TALCig}{\ensuremath{\mathtt{TAB}_{\mathcal{ALC}\iota_G}}}
\newcommand{\TALCil}{\ensuremath{\mathtt{TAB}_{\mathcal{ALC}\iota_L}}}
\newcommand{\B}{\ensuremath{\mathcal{B}}}

\newtheorem{theorem}{Theorem}
\newtheorem{definition}[theorem]{Definition}
\newtheorem{example}[theorem]{Example}
\newtheorem{lemma}[theorem]{Lemma}

\newtheorem{proposition}[theorem]{Proposition}

\newtheorem*{theorem*}{Theorem}
\newtheorem*{proposition*}{Proposition}

\title{Description Logics with Two Types of Definite Descriptions:\\
Complexity, Expressiveness, and Automated Deduction
}

\author{
    Michał Sochański\equalcontrib\textsuperscript{\rm 1}, Przemysław Andrzej Wałęga\equalcontrib\textsuperscript{\rm 1,2}, Michał Zawidzki\equalcontrib\textsuperscript{\rm 1}
}
\affiliations{
    \textsuperscript{\rm 1}Department of Logic, University of Łódź,
    Poland\\
    \textsuperscript{\rm 2}School of Electronic Engineering and Computer Science, Queen Mary University of London, 
    United Kingdom

    \{michal.sochanski,przemyslaw.walega,michal.zawidzki\}@filhist.uni.lodz.pl, p.walega@qmul.ac.uk
}

\begin{document}

\maketitle

\begin{abstract}
Definite descriptions are expressions of the form ``the unique 
$x$ satisfying property $C$,'' which allow reference to objects through their distinguishing characteristics. They play a crucial role in ontology and query languages, offering an alternative to proper names (IDs), which lack semantic content and serve merely as placeholders. 

In this paper, we introduce two extensions of the well-known description logic $\ALC$ with local and global definite descriptions, denoted $\ALCil$ and $\ALCig$, respectively. We define appropriate bisimulation notions for these logics, enabling an analysis of their expressiveness. We show that although both logics share the same tight \EXPT{} complexity bounds for concept and ontology satisfiability, $\ALCig$ is strictly more expressive than $\ALCil$. Moreover, we present tableau-based decision procedures for satisfiability in both logics, provide their implementation, and report on a series of experiments. The empirical results demonstrate the practical utility of the implementation and reveal interesting correlations between performance and structural properties of the input formulas.
\end{abstract}

\section{Introduction}

Definite descriptions (DDs)---expressions of the form ``the unique $x$ satisfying property $C$''---serve as complex terms identifying individuals via uniquely characterising properties. Their study originates in philosophical logic, where the main focus has been on semantics~\cite{russell1905denoting,PelLin2005,HilBer68,Rosser78,Lambert2001}. In recent decades, DDs have drawn renewed interest in formal logic, including classical, intuitionistic, and temporal systems~\cite{Fitting2023,Indrzejczak2023c,Indrzejczak2023a,Indrzejczak2023b,Indrzejczak2024a,Indrzejczak2021a,Indrzejczak2023e,Indrzejczak2023d,Kurbis2019a,Kurbis2019b,Kurbis2025,Orlandelli2021}. This line of research has also led to reasoning systems supporting DDs, including KeYamera X~\cite{Bohrer2019}, PROVER9~\cite{Oppenheimer2011}, and Isabelle/HOL~\cite{Benzmuller2020}.

Of particular relevance  is the application of DDs in \emph{Knowledge Representation and Reasoning} (KRR), where they enable precise identification of individuals while encoding structural constraints---a capability surpassing that of non-descriptive names such as opaque IDs~\cite{borgida2016referring2,borgida2016referring,Borgidaijcai2017,TomanTreeModel16,toman2018identity,toman2019finding,toman2019identity}. 
Within \emph{description logics} (DLs), this has motivated the introduction of DD operators into the logical language~\cite{areces2008referring,Ren2010,neuhaus2020free,toman2019finding}. In particular, \citet{artale2021free} introduced concepts $\{\iota C\}$, denoting the singleton containing the unique individual satisfying concept $C$, or the empty set if no such individual exists.
This allows for succinct representation of concepts such as:
\[
\{ \iota (\mathsf{building}\sqcap\forall\mathsf{tallThan}.\neg\mathsf{building}) \},
\]
which captures the  expression ``the tallest building.'' We refer to such constructs as \emph{local} DDs.

We contrast \emph{local} DDs with newly introduced \emph{global} DDs of the form $\iota C.D$, expressing that ``the individual satisfying $C$ also satisfies $D$.'' For example, the concept
\[
\iota (\mathsf{building}\sqcap\forall\mathsf{tallThan}.\neg\mathsf{building}).
\exists \mathsf{locIn}.\{ Dubai \}
\]
formalises the statement ``the tallest building is located in Dubai.'' 
This type of  DDs builds on recent developments in modal and hybrid logics~\cite{walega2023,Walega2024,Indrzejczak2023e}.

Despite growing interest in DDs within DLs, key foundational questions have remained open. In particular, the \emph{computational complexity}, \emph{expressive power}, and \emph{reasoning procedures} for $\ALC$ extended with local or global DDs have not been systematically studied. While the complexity analysis proves relatively straightforward, characterising expressive power is significantly more challenging. Notably, devising suitable \emph{bisimulations} for such logics is non-trivial. 
Existing bisimulations are for local DDs only and assume the presence of nominals and the universal role---features that simplify the task~\cite{artale2021free}. To the best of our knowledge, bisimulations for \ALC{} with DDs alone have not been proposed.
Moreover, no existing DL reasoning system appears to support DDs, despite their practical motivation.

This paper aims to address these gaps. Our main contributions are as follows:

\smallskip

\noindent{\scriptsize${\bullet}$}  We introduce three extensions of \ALC{} with definite descriptions: \ALCil{} (local), \ALCig{} (global), and \ALCi{} (both). These are formalised in \Cref{DLs}.\smallskip

\noindent{\scriptsize${\bullet}$} To analyse expressivity, we define bisimulations for each logic in \Cref{sec:exress}. They rely on novel, non-trivial conditions, which we show can be reduced to standard \ALC{} bisimulation checks. We also provide algorithms for this reduction.\smallskip

\noindent{\scriptsize${\bullet}$} Using these bisimulations, we study expressivity via equivalence-preserving concept translations. We show that $\ALCil$ is strictly less expressive than $\ALCig$, while $\ALCig$ and $\ALCi$ are equally expressive. All three are $\EXPT$-complete for concept and ontology satisfiability.\smallskip

\noindent{\scriptsize${\bullet}$} In \Cref{sec:tableau}, we present tableau-based decision procedures for all three logics. Despite differing semantics of the two DD types, similar tableau rules suffice to handle both.\smallskip

\noindent{\scriptsize${\bullet}$} We implement these procedures and evaluate them on custom benchmarks (\Cref{sec:implement}). Results confirm their viability and show links between structural features of DDs and reasoning performance.

\section{Description Logics with Definite Descriptions}\label{DLs}

\paragraph{Syntax}
Let $\mathsf{N_C}$, $\mathsf{N_R}$, and $\mathsf{N_I}$ be countably infinite, pairwise disjoint sets of \emph{atomic concept names}, \emph{role names}, and \emph{individual names}, respectively. 
$\ALCi$ \emph{concepts} $C$ are defined by the following grammar:
$$
C \coloneqq A \mid \neg C \mid (C \sqcap C) \mid \exists r.C \mid \{ \iota C \} \mid \iota C. C,
$$
where $A \in \mathsf{N_C}$ and $r \in \mathsf{N_R}$.
Standard abbreviations are used for other logical constructs: 
$\bot \coloneqq A \sqcap \neg A$, 
$\top \coloneqq \neg \bot$, 
$C \sqcup D \coloneqq \neg(\neg C \sqcap \neg D)$, and 
$\forall r.C \coloneqq \neg \exists r.\neg C$.
An $\ALCi$ \emph{concept inclusion} (CI) is a formula of the form $C \sqsubseteq D$, where $C$ and $D$ are $\ALCi$ concepts. 
We write $C \equiv D$ as shorthand for $C \sqsubseteq D$ and $D \sqsubseteq C$.
An  \emph{assertion} is either $a:C$ or $r:(a_1, a_2)$, where $a, a_1, a_2 \in \mathsf{N_I}$, $C$ is a concept, and $r \in \mathsf{N_R}$.
An \emph{ABox} $\mathcal{A}$ is a finite set of assertions; a \emph{TBox} $\T$ is a finite set of concept inclusions. An \emph{ontology} $\Ont$ consists of a TBox and an ABox.
Where clear from context, we refer simply to \emph{atomic concepts}, \emph{roles} and \emph{individuals} without mentioning names explicitly.

The description logic $\ALCil$ is the fragment of $\ALCi$ that includes \emph{local definite descriptions} $\{\iota C\}$ but excludes \emph{global definite descriptions}  $\iota C.D$.
Conversely, $\ALCig$ includes only global descriptions.

\paragraph{Semantics}
An \emph{interpretation} is a pair $\I = (\Delta^{\I}, \cdot^{\I})$ consisting of a non-empty domain $\Delta^{\I}$ and a function that maps:
(i) atomic concepts $A \in \mathsf{N_C}$ to subsets of $\Delta^{\I}$,
(ii) roles  $r \in \mathsf{N_R}$ to subsets of $\Delta^{\I} \times \Delta^{\I}$,
(iii) individuals $a \in \mathsf{N_I}$ to elements of $\Delta^{\I}$.
This function extends to complex concepts:
\begin{align*}
(\neg C)^{\I} & \coloneqq \Delta^{\I} \setminus C^{\I},\qquad 
 (C \sqcap D)^{\I} \coloneqq C^{\I} \cap D^{\I}, \\
(\exists r.C)^{\I} & \coloneqq \{ d \in \Delta^{\I} \mid (d, e) \in r^{\I} \text{ for some } e \in C^{\I} \}, \\
(\{ \iota C \})^{\I} & \coloneqq
\begin{cases}
\{ d \}, & \text{if } C^{\I} = \{d\} \text{ for some } d \in \Delta^{\I}, \\
\emptyset, & \text{otherwise},
\end{cases}
\\
(\iota C . D)^\I & \coloneqq
\begin{cases}
\Delta^\I, & \text{if } C^{\I} = \{d\} \subseteq D^\I \text{ for some } d \in \Delta^{\I}, \\
\emptyset, & \text{otherwise}.
\end{cases}
\end{align*}
A concept $C$ is \emph{satisfied} in $\I$ if $C^{\I} \neq \emptyset$, and \emph{satisfiable} if such $\I$  exists. A \emph{pointed interpretation} is a pair $(\I, d)$ with $d \in \Delta^\I$. We write 
$(\I, d) \equiv_{\Lan} (\J, e)$ if for every concept $C$ from a logic $\Lan$, we have $d \in C^{\I}$ iff $e \in C^{\J}$.
Satisfaction of axioms of ontology $\Ont$ in $\I$, written $\I \models \alpha$, is defined as:
\begin{align*}
\I & \models C \sqsubseteq D &&\text{iff}&& C^{\I} \subseteq D^{\I}, \\
\I & \models a:C &&\text{iff}&& a^{\I} \in C^{\I}, \\
\I & \models r:(a_1, a_2) &&\text{iff}&& (a_1^{\I}, a_2^{\I}) \in r^{\I}.
\end{align*}
Interpretation $\I$ is a \emph{model} of an ontology $\Ont$ (written $\I \models \Ont$) if $\I \models \alpha$ for all $\alpha \in \Ont$. An ontology is \emph{satisfiable} if it has a model, and a concept $C$ is \emph{satisfiable w.r.t.\ an ontology} $\Ont$ if $C$ is satisfied in a model of $\Ont$.

We compare the expressive power of description logics via equivalence-preserving translations between their concepts.
A logic $\Lan$ is \emph{not more expressive} than $\Lan'$, denoted $\Lan \leq \Lan'$, if every $\Lan$-concept has an equivalent  $\Lan'$-concept.
It is \emph{strictly less expressive} if $\Lan \leq \Lan'$ but $\Lan' \not\leq \Lan$,
and \emph{equally expressive} if both $\Lan \leq \Lan'$ and $\Lan' \leq \Lan$.


\section{Expressiveness and Complexity}\label{sec:exress}

We analyse and compare the logics $\ALCil$, $\ALCig$, and $\ALCi$. Although their expressive power differs, all three have the same computational complexity.

\begin{theorem}\label{complexity}
In $\ALCil$, $\ALCig$, and $\ALCi$, both concept and ontology satisfiability are \EXPT{}-complete.
\end{theorem}

\begin{proof}[Proof sketch]
For upper bounds, we reduce $\ALCi$ ontology satisfiability to that of $\ALCOiu$, which is  $\EXPT$-complete~\cite[Thm. 2]{artale2021free}. Local DDs $\{\iota C\}$ are allowed in $\ALCOiu$, and each global DD $\{\iota C.D\}$ is replaced with $\exists u. ( \{ \iota C \} \sqcap D)$, where $u$ is the universal role.
This polynomial reduction constructs an equivalent ontology.


For lower bounds, it suffices to show \EXPT{}-hardness for concept satisfiability in $\ALCil$ and $\ALCig$. We give logspace reductions from satisfiability of an $\ALC$ concept $C$ w.r.t.\ a TBox $\T$, known to be \EXPT{}-complete~\cite[Thm. 3.27]{baader2003description}.
For $\ALCil$, we construct $C'$ as:
\begin{center}
$C \sqcap 
\bigsqcap_{(D \sqsubseteq E) \in \T}
\left(
(\neg D \sqcup E) \sqcap 
\left\{\iota\left(\neg(\neg D \sqcup E) \sqcup A_{D \sqsubseteq E}\right)\right\}
\right)$,
\end{center}
where each $A_{D \sqsubseteq E}$ is a fresh atomic concept. Then $C'$ is satisfiable iff $C$ is satisfiable w.r.t.\ $\T$.
For $\ALCig$, we replace the above local definite descriptions with the following:
$
A_{D \sqsubseteq E} \sqcap 
\iota\left(\neg(\neg D \sqcup E) \sqcup A_{D \sqsubseteq E}\right).\top.
$
\end{proof}

Despite having the same complexity, the logics differ in expressive power. We first observe that local descriptions can be encoded using global ones:

\begin{proposition}\label{translations}
There is an exponential translation of $\ALCil$ concepts into equivalent $\ALCig$ concepts, and a polynomial translation of $\ALCil$ ontologies into conservative extensions in $\ALCig$.
\end{proposition}

\begin{proof}[Proof sketch]
The exponential translation replaces $\{\iota C\}$ with $C \sqcap \iota C.\top$.
The polynomial version replaces $\{\iota C\}$ with $A_C \sqcap \iota A_C.\top$ and adds axioms $A_C \equiv C$.
\end{proof}
We conclude that
$\ALCil \leq \ALCig = \ALCi.$ 
Next, we will introduce bisimulations and show that $\ALCil < \ALCig$.
For this, we will exploit the following notion of \emph{names} of individuals in an interpretation:

\begin{definition}
Let $\Delta' \subseteq \Delta^\I$. The set $\names(\Delta', \I)$ consists of all $\ALC$ concepts $C$, with $C^\I = \{d\}$ for some $d \in \Delta'$.
\end{definition}

\begin{example}\label{example}
Let $\Delta^\I = \{a,b\}$ with $A^\I = \emptyset$, and $\Delta^\J = \{c,d,e\}$ with $A^\J = \{e\}$ (see \Cref{fig:example}).
Then $\names(\{a,b\}, \I) = \names(\{c,d\}, \J) = \emptyset$, since no $\ALC$ concept uniquely identifies an individual in either set.
However, $\names(\{c,d,e\}, \J)$ is non-empty, as it contains for example  $A$, $\neg\neg A$, and $A \sqcup \exists r.\top$.
\end{example}

\begin{figure}[t]
\centering
\begin{tikzpicture}
\tikzset{>=latex}

\node[draw,circle,minimum size=18pt] (w1) at (0,0) {a};
\node[draw,circle,minimum size=18pt] (w2) at (1,0) {b};

\node[draw,circle,minimum size=18pt] (w3) at (4,0) {c};
\node[draw,circle,minimum size=18pt](w4) at (5,0) {d};
\node[draw,circle,minimum size=18pt] (w5) at (6,0) {e};
\node[draw=none,below=0.1 of w5] (t5) {$A$};

\node[draw=none] at (0.5,0.8) {$\I$};
\node[draw=none] at (5,0.8) {$\J$};

\draw [-, dashed, blue] (w2) --  node[above] {$Z$} (w3);
\draw [-, dashed, blue] (w2) to [out=30,in=-180-30]   (w4);
\draw [-, dashed, blue] (w1) to [out=-40,in=-180+40]   (w4);
\draw [-, dashed, blue] (w1) to [out=-30,in=-180+30]   (w3);
\end{tikzpicture}
\caption{Maximal $\ALCil$ bisimulation between $\I$ and $\J$}\label{fig:example}
\end{figure}

We now define bisimulations for $\ALCil$ and $\ALCig$. Note that by \Cref{translations}, a separate definition for $\ALCi$ is unnecessary.

\begin{definition}\label{def::bisimulations}
An \ALCil{} bisimulation between interpretations $\I$ and $\J$ is a relation $Z \subseteq \Delta^\I \times \Delta^\J$ such that $Z = \emptyset$ or, for all $(d,e) \in Z$, every  atomic concept  $A$, and  role $r$:
\begin{description}
\item[Atom] $d \in A^\I$ iff $e \in A^\J$,
\item[Forth] if $(d,d') \in r^\I$, then there exists $ e'$ such that $(e,e') \in r^\J$ and $(d',e') \in Z$,
\item[Back] if $(e,e') \in r^\J$, then there exists $d'$ such that $(d,d') \in r^\I$ and $(d',e') \in Z$,
\item[Names$_L$] $\names(\Dom(Z), \I) = \names(\Rng(Z), \J)$.
\end{description}
An \ALCig{} bisimulation is defined identically, except that \textbf{Names}$_L$ is replaced with:
\begin{description}
\item[Names$_G$] $\names(\Delta^\I, \I) = \names(\Delta^\J, \J)$.
\end{description}
We write $(\I, d) \sim_{\Lan} (\J, e)$ if $(d,e) \in Z$ for some $\Lan$-bisimulation $Z$, where $\Lan \in \{\ALCil, \ALCig\}$.
\end{definition}

\Cref{fig:example}, presents the maximal $\ALCil$ bisimulation $Z$ between $\I$ and $\J$ It satisfies \textbf{Names}$_L$, as $\names(\Delta^\I, \I) = \names(\Delta^\J, \J) = \emptyset$. However, there is no $\ALCig$ bisimulation betwee $\I$ and $\J$, as \textbf{Names}$_G$ fails: $A \in \names(\Delta^\J,\J)$ but $A \notin \names(\Delta^\I,\I)$.

It is worth observing that
our bisimulations differ from those used for $\ALCOiu$ by \citet{artale2021free}, which rely on totality and ``counting up to one'' via the universal role and nominals. These conditions are too strong for $\ALCil$, as illustrated in \Cref{fig:example}.
While our \textbf{Names}$_L$ and \textbf{Names}$_G$ conditions are non-standard and appear to require quantification over all concepts, we show in Algorithms 1 and 2 how they can be verified procedurally. Before that, we prove that our bisimulations preserve concept satisfiability, and that the converse holds for $\omega$-saturated\footnote{See, e.g., \citet{DBLP:books/daglib/0067423} for the definition.} interpretations.



\begin{theorem}\label{bisimulation_thm}
For all pointed  interpretations $(\mathcal{I}, d)$ and $(\mathcal{J}, e)$, and both $\Lan \in \{ \ALCil, \ALCig \}$ the following hold:
\begin{enumerate}
\item if $(\I, d) \sim_{\Lan} (\J, e)$, then $(\I, d) \equiv_{\Lan} (\J, e)$,
\item if $(\I, d) \equiv_{\Lan} (\J, e)$ and $\I, \J$ are $\omega$-saturated, then $(\I, d) \sim_{\Lan} (\J, e)$.
\end{enumerate}
\end{theorem}
\begin{proof}[Proof sketch]
We first prove Statement~1 by induction on the structure of $\Lan$-concepts $C$, showing $d \in C^\I$ if and only if $e \in C^\J$.
If $C$ is atomic, or of the form $\neg D$, $D \sqcap E$, or $\exists r.D$, the result follows from conditions \textbf{Atom}, \textbf{Forth}, and \textbf{Back}, as in the standard $\ALC$ case~\cite{baader2017introduction}.
If $C = \{\iota D\}$ and $d \in C^\I$, then $D^\I = \{d\}$, so $D \in \names(\Dom(Z), \I)$. By the inductive hypothesis, $e \in D^\J$, and by \textbf{Names}$_L$, $D \in \names(\Rng(Z), \J)$, hence $D^\J = \{e\}$, and $e \in (\{\iota D\})^\J$. The converse direction is analogous.
If $C = \iota D.E$ and $d \in C^\I$, then $\{D, D \sqcap E\} \subseteq \names(\Delta^\I, \I)$. By \textbf{Names}$_G$, these concepts are also in $\names(\Delta^\J, \J)$, implying $e \in (\iota D.E)^\J$. The converse again is similar.

For Statement~2, we begin with $\Lan = \ALCig$. Since $(\I, d) \equiv_{\ALCig} (\J, e)$ implies $(\I, d) \equiv_{\ALC} (\J, e)$, and $\I$, $\J$ are $\omega$-saturated, standard results for $\ALC$~\cite{baader2017introduction} imply that there exists an $\ALC$-bisimulation $Z$ with $(d,e) \in Z$.
To show that $Z$ is an $\ALCig$-bisimulation, it remains to prove \textbf{Names}$_G$. If $C \in \names(\Delta^\I, \I)$, then $d \in (\iota C.\top)^\I$, hence $e \in (\iota C.\top)^\J$, so $C \in \names(\Delta^\J, \J)$. The reverse direction is symmetric.
For $\Lan = \ALCil$, we define 
$Z := \{(k, l) \in \Delta^\I \times \Delta^\J \mid (\I, k) \equiv_{\ALCil} (\J, l)\}$,
so $(d, e) \in Z$. Conditions \textbf{Atom}, \textbf{Forth}, and \textbf{Back} hold by standard arguments for $\ALC$ bisimulations, extended with a translation of $\ALCil$ into first-order logic.
To show \textbf{Names}$_L$, assume $C \in \names(\Dom(Z), \I)$. Then some $d' \in \Dom(Z)$ satisfies $d' \in (\{\iota C\})^\I$. Since $(d', e') \in Z$ for some $e' \in \Delta^\J$, we conclude $e' \in (\{\iota C\})^\J$, so $C \in \names(\Rng(Z), \J)$. Thus, $\names(\Dom(Z), \I) \subseteq \names(\Rng(Z), \J)$. The reverse inclusion follows analogously, proving equality.
\end{proof}

We now apply our bisimulations to show that $\ALCil < \ALCig$. Combined with \Cref{translations}, this yields the following expressiveness result.

\begin{theorem}
The following expressive power relations hold:
$\ALCil < \ALCig = \ALCi$.
\end{theorem}

\begin{proof}
By \Cref{translations}, it suffices to show $\ALCig \not\leq \ALCil$.
Assume, for contradiction, that $\iota A.\top$ is equivalent to some $\ALCil$ concept $C$.
Consider the interpretations $\I$ and $\J$ from \Cref{example}, with the maximal $\ALCil$ bisimulation $Z$ shown in \Cref{fig:example}. Since $(\I, a) \sim_{\ALCil} (\J, c)$, \Cref{bisimulation_thm} gives $a \in C^\I$ iff $c \in C^\J$.
However, by construction, $\I \not\models \iota A.\top(a)$ and $\J \models \iota A.\top(c)$, i.e., $a \notin C^\I$ and $c \in C^\J$---a contradiction.
\end{proof}

While our bisimulations accurately characterise concept equivalence, we have not yet addressed how to decide bisimilarity between two pointed interpretations. We now present an algorithm for computing the maximal bisimulation between two finite interpretations, identifying all bisimilar pairs.
The following notion plays a central role:

\begin{definition}
Let $\Delta' \subseteq \Delta^\I$. The set $\named(\Delta', \I)$ of \emph{named individuals} comprises all $d \in \Delta'$ such that $d \in C^\I$ for some $C \in \names(\Delta', \I)$.
\end{definition}

For instance, in the interpretations $\I$ and $\J$ from \Cref{example}, we have $\named(\Delta^\I, \I) = \emptyset$ and $\named(\Delta^\J, \J) = \{e\}$.

In contrast to computing names, identifying named individuals is straightforward: they are exactly those not $\ALC$-bisimilar to any other individual in the interpretation.
The next theorem links named individuals to names, showing that once the named individuals are known, one can determine whether two interpretations have the same names. Below, we call relation $Z$  \emph{total} if it is both left- and right-total.

\begin{theorem}\label{thm:total}
Let $\I$ and $\J$ be  finite interpretations and let $\Delta' \subseteq \Delta^\I$ and $\Delta'' \subseteq \Delta^\J$.
If $\names(\Delta',\I) \neq \emptyset \neq \names(\Delta'',\J)$, then $\names(\Delta',\I) = \names(\Delta'',\J)$ if and only if 
the maximal \ALC{}-bisimulation $Z$ between $\I$ and $\J$ is a total relation and the restriction of $Z$ to $\named(\Delta',\I) \times \named(\Delta'',\J)$ is also total.
%
%
\end{theorem}
\begin{proof}[Proof sketch]
Assume that $\names(\Delta', \I) = \names(\Delta'', \J) \neq \emptyset$. To show that the restriction of $Z$ to $\named(\Delta', \I) \times \named(\Delta'', \J)$ is total, suppose for contradiction that some $d \in \named(\Delta', \I)$ has no $Z$-related counterpart in $\named(\Delta'', \J)$. Then there exists $C \in \names(\Delta', \I)$ such that $C^\I = \{d\}$ and $C^\J = \{e\}$ for some $e \in \Delta''$. Since $(d, e) \notin Z$, there exists an $\ALC$ concept $D$ with $d \in D^\I$ but $e \notin D^\J$, so $C \sqcap D \in \names(\Delta', \I)$ but $C \sqcap D \notin \names(\Delta'', \J)$---a contradiction. To show that $Z$ is total, suppose some $d \in \Delta^\I \setminus \named(\Delta', \I)$ has no $Z$-related individual in $\Delta^\J$. As $\names(\Delta'', \J) \neq \emptyset$, there exists $e^* \in \named(\Delta'', \J)$ with $C^\J = \{e^*\}$ for some $C$. For each $e \in \Delta^\J \setminus \{e^*\}$, choose $C_e$ with $e \in C_e^\J$ and $d \notin C_e^\I$, and let $D = C \sqcup \bigsqcap_{e \neq e^*} \neg C_e$. Then $D^\J = \{e^*\}$, so $D \in \names(\Delta'', \J) = \names(\Delta', \I)$, and since $d \in D^\I$, we get $d \in \named(\Delta', \I)$---a contradiction.

Assume $Z$ and its restriction are total. Suppose, for contradiction, that $C \in \names(\Delta', \I)$ but $C \notin \names(\Delta'', \J)$. Then $C^\I = \{d\}$ for some $d \in \named(\Delta', \I)$, and there exists $e \in \named(\Delta'', \J)$ with $(d, e) \in Z$ and $e \in C^\J$. Since $C \notin \names(\Delta'', \J)$, there is $e' \neq e$ with $e' \in C^\J$. By totality, there exists $d' \in \Delta^\I$ with $(d', e') \in Z$. If $d' \neq d$, then $\{d, d'\} \subseteq C^\I$ contradicts uniqueness. If $d' =d$, and for $D \in \names(\Delta'', \J)$, $D^\J = \{e\}$, then $d \in D^\I$ but $e \notin D^\J$, so $(d, e') \in Z$ contradicts that $Z$ is a bisimulation.
\end{proof}

In \Cref{alg:maxbisimALCil}, we use \Cref{thm:total} to compute the maximal $\ALCi$ bisimulation between finite interpretations $\I$ and $\J$.
The algorithm first computes three maximal $\ALC$ bisimulations: $Z$, $Z_\I$, and $Z_\J$, which are used to determine the sets $N_\I = \named(\Delta^\I, \I)$ and $N_\J = \named(\Delta^\J, \J)$, as well as $N_\I^Z = \named(\Dom(Z), \I)$ and $N_\J^Z = \named(\Rng(Z), \J)$.
If both $N_\I^Z$ and $N_\J^Z$ are empty, then \textup{\textbf{Names}}$_L$ holds and the algorithm returns $Z$. If only one is empty, then \textup{\textbf{Names}}$_L$ fails, and the algorithm returns $\emptyset$.
Otherwise, \Cref{thm:total} provides an equivalent condition for \textup{\textbf{Names}}$_L$: if it holds, the algorithm returns $Z$; otherwise, it returns $\emptyset$.
For example, when applied to $\I$ and $\J$ from \Cref{fig:example}, the algorithm returns the bisimulation $Z$ shown therein. 

\begin{algorithm}[ht]
\SetKwFor{Loop}{loop}{}{}
\SetKwComment{Comment}{$//$ }{}
\SetKwInput{Input}{Input}
\SetKwInput{Output}{Output}
\Input{interpretations $\I$ and $\J$}
\Output{maximal \ALCil{} bisimulation for $\I$ and $\J$}

$Z \coloneqq \maxB(\I,\J)$; \label{MLDD1}

$Z_\I \coloneqq \maxB(\I,\I)$; \label{MLDD2}

$Z_\J \coloneqq \maxB(\J,\J)$; \label{MLDD3}

$N_\I \coloneqq \{ d \in \Delta^\I  \mid (d,e) \not\in Z_\I \text{ for all } e \neq d \text{ in } \Delta^\I \}$; \label{MLDD4}

$N_\J \coloneqq \! \{ d \in \Delta^\J \! \mid \! (d,e) \not\in Z_\J \text{ for all } e \neq d \text{ in } \Delta^\J \}$; \label{MLDD5}

$N_\I^Z \coloneqq N_\I \cap \Dom(Z)$;

$N_\J^Z \coloneqq N_\J \cap \Rng(Z)$;

\lIf{$N_\I^Z= \emptyset = N_\J^Z$}{\Return $Z$}\label{MLDD6}

\lIf{$N_\I^Z = \emptyset \neq N_\J^Z$ or $N_\I^Z \neq \emptyset = N_\J^Z$}{\Return $\emptyset$}\label{MLDD65}

\lIf{$Z$ is total  over $\Delta^\I \times \Delta^\J$ and $Z \cap  (N_\I^Z \times N_\J^Z)$ is total  over $N_\I^Z \times N_\J^Z$}{\Return $Z$}\label{MLDD7}
\lElse{{\Return $\emptyset$}}\label{MLDD8}

\caption{Maximal \ALCil{}-bisimulation}
\label{alg:maxbisimALCil}
\end{algorithm}

To compute the maximal $\ALCi$ bisimulation, we introduce Algorithm~2 which modifies Algorithm~\ref{alg:maxbisimALCil} by replacing Lines~6 and~7 with the following:
$$
{\tiny \textbf 6} \; N_\I^Z \coloneqq N_\I ;
\qquad \quad 
{\tiny \textbf 7} \; N_\J^Z   \;\coloneqq N_\J ;
$$
%
%
We can show that both algorithms are correct:


\begin{theorem}\label{correctAlg}
Algorithms~\ref{alg:maxbisimALCil} and~2 return, respectively, the maximal \ALCil{} and \ALCi{} bisimulations for $\I$ and $\J$. 
\end{theorem}

\begin{proof}
The argumentation exploits  \Cref{thm:total}, which provides the necessary condition for the bisimulation checks. What remains is to justify that $N_\I = \named(\Delta^\I, \I)$ and $N_\J = \named(\Delta^\J, \J)$.
This follows from the Hennessy-Milner property for \ALC{} bisimulations and the finiteness of $\I$ and $\J$. Specifically, if $d \in N_\I$, then for every $e \neq d$, there exists an \ALC{} concept $C_e$ such that $d \in C_e^\I$ and $e \notin C_e^\I$. Let $D = \bigsqcap_{e\neq d} C_e$; then $D^\I = \{d\}$, so $d \in \named(\Delta^\I, \I)$.
Conversely, if $d \notin N_\I$, then there exists $e \neq d$ such that $(\I, d) \sim_{\ALC} (\I, e)$, and thus $d$ satisfies exactly the same \ALC{} concepts as $e$, implying $d \notin \named(\Delta^\I, \I)$. The argument for $N_\J = \named(\Delta^\J, \J)$ is analogous.
\end{proof}

\section{Tableaux Systems}\label{sec:tableau}

\begin{figure}[t]
\centering 
\begin{scriptsize}

\textbf{ABox rules:}\hfill\medskip

$(ABox_\mathsf{I})$\ $\dfrac{a:C\in ABox}{a:C}$\qquad $(ABox_r)$\ $\dfrac{r(a,a')\in ABox}{r(a,a')}$\medskip

\begin{minipage}[t]{.54\linewidth}
\centering
\textbf{TBox rule:}\hfill\medskip

$(TBox)$\ $\dfrac{C\sqsubseteq D\in TBox,a:E}{a:\neg(C\sqcap\neg D)}$
\end{minipage}\hfill
\begin{minipage}[t]{.44\linewidth}
\centering
\textbf{Clash rule:}\hfill\medskip

    $(\bot)\ \dfrac{a:C,a:\neg C}{\bot}$
\end{minipage}\medskip

\textbf{Propositional rules:}\hfill\medskip

    $(\neg\neg)$\ $\dfrac{a:\neg\neg C}{a: C}$\qquad \ $(\sqcap)$\ $\dfrac{a:C\sqcap D}{a:C,a:D}$\qquad
    $(\neg\sqcap)$\ \ $\dfrac{a:\neg(C\sqcap D)}{a:\neg C\mid a:\neg D}$\medskip

\textbf{Role rules:}\hfill\medskip

    $(\exists r)$\ $\dfrac{a:\exists r.C}{b:C,r:(a,b)}$\qquad
    $(\neg\exists r)$\ $\dfrac{a:\neg\exists r.C,r:(a,a')}{a':\neg C}$\medskip

\textbf{Global definite description rules:}\hfill\medskip

    $(\iota^g_1)$\ $\dfrac{a:\iota C.D}{b:C,b:D}$\qquad
    $(\iota^g_2)$\ $\dfrac{a:\iota C.D,a':C,a'':C,a':E}{a'':E}$\\[5pt]
    $(\neg\iota^g)$\ $\dfrac{a:\neg\iota C.D,a':E}{a':\neg C\mid a':\neg D\mid \parbox[t]{1.9cm}{\centering$b:C,b:A_C^g$,\\$b':C,b':\neg A_C^g$}}$\qquad
    $(cut^g_\iota)$\ $\dfrac{a:\iota C.D,a':E}{a':C \mid a':\neg C}$\medskip

\textbf{Local definite description rules:}\hfill\medskip

    $(\iota^\ell_1)$\ $\dfrac{a:\{\iota C\}}{a:C}$\qquad
    $(\iota^\ell_2)$\ $\dfrac{a:\{\iota C\},a':C,a'':C,a':D}{a'':D}$\\[5pt]
    $(\neg\iota^\ell)$\ $\dfrac{a:\neg\{\iota C\}}{a:\neg C\mid \parbox[t]{1.8cm}{\centering$a:\neg A_C$,\\$b:C,b: A_C$}}$\qquad
    $(cut^\ell_\iota)$\ $\dfrac{a:\{\iota C\},a':D}{a':C\mid a':\neg C}$
    \end{scriptsize}\smallskip

    \hrule\smallskip
    \begin{scriptsize}
    $^*$ $b,b'$ occurring in the conclusion of a rule are fresh.\hfill
    \end{scriptsize}
\caption{Rules of $\TALCi$}
\label{fig::TableauRules}
\end{figure}

In this section, we present tableau calculi 
$\TALCil$, $\TALCig$, and $\TALCi$, for all three logics.
The first two share rules for common constructs, but differ in handling local and global definite descriptions, whereas $\TALCi$ combines these systems.

\subsubsection*{Rules}
Given a concept $C$ (optionally with ontology $\mathcal{O}$), our calculi return \texttt{sat} if $C$ is satisfiable (w.r.t.\ $\mathcal{O}$), and \texttt{unsat} otherwise. 
Rules are applied to assertions to incrementally build a tableau tree.
The root contains $a:C$, where $a$ is fresh and does not occur in $\mathcal{O}$. A \emph{branch} $\B$ is a path from root to leaf; if $\B'$ extends $\B$, we write $\B \subseteq \B'$. If it does not lead to confusion, we treat branches as sets of assertions.
For an individual $a$ and branch $\B$, the \emph{theory} of $a$, written $\theory{\B}{a}$, is $\{C \mid a:C \in \B\}$; we say $a$ \emph{satisfies} $C$ on $\B$ if $a:C \in \B$. A rule has the form $\frac{\mathcal{P}r}{\mathcal{C}on_1 \mid \cdots \mid \mathcal{C}on_m}$, with $m$ the branching factor. Rules are \emph{deterministic} if $m=1$, and \emph{branching} otherwise. The calculus’s branching factor is the largest $m$ among its rules.
%
A rule $(\mathsf{r})$ applies to premise $\mathcal{P}r$ if (i) it is not \emph{blocked}, (ii) no conclusion already appears on the branch, and (iii) the branch is \emph{open}. A branch is \emph{closed} if a clash occurs; otherwise, it is open. It is \emph{saturated} if open and no rules are applicable.

Figure~\ref{fig::TableauRules} lists the rules of $\TALCi$. Those for standard $\ALC$ constructs are omitted here. We highlight the handling of DDs and the blocking condition for $(\exists r)$.\smallskip

\noindent\textit{Blocking.} Rule $(\exists r)$ is blocked for $a:\exists r.C$ if:
\begin{itemize}
\item[($block_\exists$)] There exists an individual $a'$ on $\B$ such that $a':C \in \B$ and $a':\neg D \in \B$ for all $D$ with $a:\neg\exists r.D \in \B$.
\end{itemize}
In that case, $a'$ serves as a proxy $r$-successor of $a$. Blocking can be lifted if new $a:\neg\exists r. E$ assertions are added such that $a':E \notin \B$. This is known as pattern-based blocking~\cite{Kaminski2009}.\smallskip

\noindent\textit{Global DDs.} Rule $(\iota_1^g)$ introduces a fresh individual satisfying both $C$ and $D$, unless an existing one already satisfies $C$, in which case only $D$ is propagated (to the least such individual w.r.t.\ lexicographic order, if needed). Rule $(\iota_2^g)$ ensures uniqueness by merging the theories of any two individuals satisfying $C$. Rule $(\neg\iota^g)$ enforces non-uniqueness: for each individual $a'$, at least one of the following holds: (i) $a':\neg C$, (ii) $a':\neg D$, or (iii) two distinct individuals satisfy $C$. In the third case, further applications to $a':\neg\iota C.E$ are blocked. Rule $(cut_\iota^g)$ ensures decisiveness: every individual satisfies either $C$ or $\neg C$ for any positive DD $\iota C.D$. The rule is crucial for the completeness of $\TALCi$. For example, the tableau for the unsatisfiable concept $(\iota \neg(C \sqcap D).\top) \sqcap (\iota C.\neg D) \sqcap (\iota D.\neg C)$ fails to close without $(cut_\iota^g)$.
 \smallskip

\noindent\textit{Local DDs.} Local DDs are handled similarly, except for negation. While $\neg\iota C.D$ enforces non-uniqueness globally, $\neg{\iota C}$ does so locally, i.e., relative to the current individual. Rule $(\neg\iota^\ell)$ ensures either the current individual does not satisfy $C$ or some other individual does. The right branch does not introduce new individuals when $(\neg\iota^\ell)$ is applied to $a':\neg\{\iota C\}$---only $a':\neg A_C$ is added in the right conclusion. In both negated-DD rules, $A_C^g$ or $A_C$ is a fresh atom (not in the input) determined by $C$.\smallskip

\noindent\textit{Priorities and Confluence.} Rules are applied in fixed order: $(\bot)$ has the highest, $(\exists)$ the lowest priority. As formulas are never removed, the calculus is \emph{cumulative}; all rules are \emph{invertible}, ensuring \emph{confluence}: rule order may affect derivation length but not outcome.

\smallskip 

\subsubsection*{Correctness} 
We prove that $\TALCi$---and so also $\TALCig$ and $\TALCil$---is sound, complete, and terminating. Recall that $\TALCi$ is \emph{sound} if, for any $\ALCi$-concept $C$ (and ontology $\mathcal{O}$), it returns \texttt{unsat} only if $C$ is unsatisfiable (w.r.t.\ $\mathcal{O}$). It is \emph{complete} if, whenever it returns \texttt{sat}, $C$ is indeed satisfiable (w.r.t.\ $\mathcal{O}$).
To establish soundness, we  show that  rules of $\TALCi$ \emph{preserve satisfiability}:

\begin{lemma}\label{lem::UnsatisfiabilityOfU}
Let $(\mathsf{r})$ be a rule of $\TALCi$ applied to a branch $\B$, and let $\B_1,\ldots,\B_n \supseteq \B$ be the resulting branches. If $\B$ is satisfiable, then so is some $\B_i$ for $i \in \{1,\ldots,n\}$.
\end{lemma}

\begin{proof}[Proof sketch]
The proof proceeds by case analysis on the rules. We illustrate it using the most complex case, $(\neg\iota^g)$.
Let $\mathcal{P}r = \{a:\neg\iota C.D,\, a':E\}$ be satisfiable. Then $a \in (\neg\iota C.D)^\I$ and $a' \in E^\I$, meaning no unique individual satisfies both $C$ and $D$. We have three cases: (i) $a' \notin C^\I$, so $a' \in (\neg C)^\I$ and $\mathcal{P}r \cup \{a':\neg C\}$ is satisfiable. (ii) $a' \notin D^\I$, so $a' \in (\neg D)^\I$ and $\mathcal{P}r \cup \{a':\neg D\}$ is satisfiable. (iii) $a' \in C^\I \cap D^\I$, and there exists $b' \in \Delta^\I$, $b' \neq a'$, such that $b' \in C^\I$. Extend $\I$ to $\I'$ so that $(A_C^g)^{\I'} = \{a'\}$. Then $a' \in (A_C^g)^{\I'}, \quad b' \in (\neg A_C^g)^{\I'}$, hence $\mathcal{P}r \cup \{b:C,\, b:A_C^g,\, b':C,\, b':\neg A_C^g\}$ is satisfiable.
\end{proof}


\begin{theorem}
For any $\ALCi$-concept $C$ (and ontology $\mathcal{O}$), if $\TALCi$ returns \texttt{unsat}, then $C$ is unsatisfiable (w.r.t.\ $\mathcal{O}$).
\end{theorem}

\begin{proof}
If $\TALCi$ returns \texttt{unsat}, then all branches of the constructed tableau are closed---i.e., the clash rule has been applied to each, indicating unsatisfiability. Since every assertion in the tableau (except the root) results from rule applications, the contrapositive of Lemma~\ref{lem::UnsatisfiabilityOfU} ensures that unsatisfiability propagates upward through the tableau, ultimately reaching $a:C$ at the root. This implies that $C$ is unsatisfiable (w.r.t.\ $\mathcal{O}$ if ABox or TBox rules were applied).
\end{proof}

For completeness, we show that if $\TALCi$ constructs a tableau with an open and saturated branch $\B$ for input concept $C$ (and ontology $\mathcal{O}$), then $C$ is satisfiable (w.r.t.\ $\mathcal{O}$), as $\B$ provides sufficient information to construct a model $\I_\B = (\Delta^{\I_\B}, \cdot^{\I_\B})$ of $C$ (and $\mathcal{O}$).

Let $\mathsf{DD}(\B) = \{C \mid a : \iota C.D \in \B \text{ or } a : {\iota C} \in \B\}$ denote the set of concepts that must have singleton extensions in $\I_\B$. For each individual $a$ on $\B$, let the \emph{representative of $a$}, $rep(a)$, be the least (w.r.t.\ lexicographic order) individual $a'$ such that both $a$ and $a'$ satisfy some $C \in \mathsf{DD}(\B)$ on $\B$, or $a$ itself if no such $C$ exists.
If the application of $(\exists r)$ to an assertion $a : \exists r . C \in \B$ was blocked (and never unblocked), then any individual $a'$ on $\B$ satisfying $\{C\} \cup \{\neg D \mid a : \neg \exists r . D \in \B\} \subseteq \theory{\B}{a'}$ is called an \emph{$(r, D)$-proxy successor} of $a$.
We define $\I_\B = (\Delta^{\I_\B}, \cdot^{\I_\B})$ as follows:
\begin{itemize}
\item $\Delta^{\I_\B} = \{rep(a) \mid a \text{ occurs on } \B\}$,
\item $a^{\I_\B} = rep(a)$ for each individual $a$ on $\B$,
\item $C^{\I_\B} = \{rep(a) \mid a : C \in \B\}$ for each concept $C$ on $\B$,
\item $r^{\I_\B} = \{(rep(a), rep(a')) \mid r : (a, a') \in \B\text{ or }a'\text{ is an }\linebreak (r,D)\text{-proxy successor of } a\text{ for some $D$}\}$ for each role $r$.
\end{itemize}
\begin{lemma}\label{lem::TruthLemma}
For any assertion $a : C \in \B$, $rep(a) \in C^{\I_\B}$.
\end{lemma}

\begin{proof}[Proof sketch]
By structural induction on $C$. We illustrate the case $C = \exists r . D$. Assume $a : \exists r . D \in \B$. Since $\B$ is saturated, two cases arise:
(i) If $(\exists r)$ was applied, then $\B$ contains $b : D$ and $r : (a, b)$. Thus $(rep(a), rep(b)) \in r^{\I_\B}$, and by the induction hypothesis, $rep(b) \in D^{\I_\B}$, hence $rep(a) \in (\exists r . D)^{\I_\B}$.
(ii) If $(\exists r)$ was blocked, then ($block_\exists$) must be satisfied. Then there exists an $(r,D)$-proxy successor of $a$ on $\B$, say $a'$. By definition, $a':D\in \B$. Thus $(rep(a), rep(a')) \in r^{\I_\B}$ and by the induction hypothesis, $rep(a') \in D^{\I_\B}$, so $rep(a) \in (\exists r . D)^{\I_\B}$.
\end{proof}

\begin{theorem}\label{thm::Completeness}
If $\TALCi$ returns \texttt{sat} for input $C$ (and $\mathcal{O}$), then $C$ is satisfiable (w.r.t.\ $\mathcal{O}$).
\end{theorem}

\begin{proof}
A \texttt{sat} output implies the existence of an open, saturated branch $\B$. By \Cref{lem::TruthLemma}, the constructed interpretation $\I_\B$ satisfies all $a : D \in \B$. Since $a : C \in \B$ for some $a$, we have $rep(a) \in C^{\I_\B}$, so $C$ is satisfiable.
If $\mathcal{O}$ is part of the input, saturation ensures all ABox rules were applied, so $\I_\B$ satisfies all ABox assertions. For TBox axioms $C \sqsubseteq D$, exhaustive application of ($TBox$) and Lemma~\ref{lem::TruthLemma} assure that $rep(a) \in C^{\I_\B}$ implies $rep(a) \in D^{\I_\B}$ for each individual $a$ on $\B$. Hence, $\I_\B$ satisfies $\mathcal{O}$, and $C$ is satisfiable w.r.t.\ $\mathcal{O}$.
\end{proof}

To prove that $\TALCi$ terminates, we show that for any input concept $C$ (and ontology $\mathcal{O}$), the tableau $\T$ constructed by applying the $\TALCi$ rules is finite. Each concept appearing in $\T$ is of one of the forms $D$, $\neg D$, $A_E^g$, $\neg A_E^g$, $A_F$, or $\neg A_F$, where $D$, $\iota E.G$ (for some concept $G$), and ${\iota F}$ are subconcepts of $C$ or of concepts in $\mathcal{O}$. Hence, the number of distinct concepts in $\T$ is bounded by $4 \cdot |C|$ (or $4 \cdot (|C| + |\mathcal{O}|)$), where $|C|$ denotes the number of symbols in $C$ (excluding parentheses), and $|\mathcal{O}|$ the total number of symbols in the concepts from $\mathcal{O}$ (also excluding parentheses). Consequently, each individual on a branch $\B$ of $\T$ can satisfy at most $4 \cdot |C|$ (or $4 \cdot (|C| + |\mathcal{O}|)$) distinct concepts.

\begin{lemma}\label{lem::NumberofPrefixesInBranch}
Let $\T$ be a tableau for input $C$ (and $\mathcal{O}$), and let $\B$ be a branch of $\T$. Then the number of distinct individuals on $\B$ is bounded by $2^{4\cdot|C|}$ (or $2^{4\cdot(|C|+|\mathcal{O}|)} + k$, where $k$ is the number of individuals occurring in the ABox).
\end{lemma}

\begin{proof}[Proof sketch]
New individuals may be introduced by the rules $(ABox_\mathsf{I})$, $(ABox_r)$, $(\exists r)$, $(\iota_1^g)$, $(\neg\iota^g)$, and $(\neg\iota^\ell)$. Every non-root individual must be introduced by one of these rules. Define a function $f$ that maps each individual $a$ not introduced by any of the ABox rules to a set of concepts:
\begin{itemize}
\item $\{D\}$, if $a$ is the root individual;
\item $\{D, E\}$, if $a$ is introduced by $(\iota_1^g)$ on $a':\iota D.E$;
\item $\{D, \pm A_D^g\}$, if introduced by $(\neg\iota^g)$ on $a':\neg\iota D.E$ and $a:\pm A_D^g\in\B$;
\item $\{D, A_D\}$, if introduced by $(\neg\iota^\ell)$ on $a':\neg\{\iota D\}$;
\item $\{D\} \cup \{\neg E \mid a':\neg\exists r.E \in \B\text{ when $a$ was introduced}\}$, if introduced by $(\exists r)$ on $a':\exists r.D$;
\end{itemize}
\noindent where $\pm A_E^g$ denotes either $A_E^g$ or $\neg A_E^g$. By construction, $f(a) \subseteq \theory{\B}{a}$. Thus, the range of $f$ is a subset of the power set of the set of concepts on $\B$, bounded by $2^{4\cdot|C|}$ (or $2^{4\cdot(|C| + |\mathcal{O}|)}$). We can show that $f$ is injective, so the bound on the number of non-ABox individuals follows. Adding the $k$ individuals from the ABox yields the claimed result.
\end{proof}
The bound in Lemma~\ref{lem::NumberofPrefixesInBranch} allows us to prove termination:

\begin{theorem}
The tableau system $\TALCi$ is terminating.
\end{theorem}

\begin{proof}
Let $\T$ be the tableau for  $C$ (and $\mathcal{O}$). Each branch $\B$ of $\T$ has at most $2^{4\cdot|C|}$ (or $2^{4\cdot(|C|+|\mathcal{O}|)} + k$, for $k$ the number of individuals in  ABox) individuals, each satisfying at most $4\cdot|C|$ (or $4\cdot(|C| + |\mathcal{O}|)$) concepts. Each rule application (except $(ABox_r)$) adds at least one concept to some individual’s theory. $(ABox_r)$ applies at most as many times as there are role assertions in the ABox, say $n$. Thus the length of $\B$ is at most $4\cdot|C|\cdot2^{4\cdot|C|}$ (or $4\cdot(|C|+|\mathcal{O}|)\cdot(2^{4\cdot(|C|+|\mathcal{O}|)}+k)+n$). As the branching factor of $\TALCi$ is also finite, $\T$ is finite, so $\TALCi$ terminates.
\end{proof}

We conclude this section with a corollary following from completeness and termination of $\TALCi$:

\begin{theorem}
Let $C$ be a satisfiable $\ALCi$-concept. There exists a model $\I$ of $C$ of size at most exponential in $|C|$.
\end{theorem}

\begin{proof}
Let $\T$ be a tableau for input $C$. Since $\TALCi$ is complete and $C$ is satisfiable, $\T$ contains an open branch $\B$. Build an interpretation $\I_\B=(\Delta^{\I_\B},\cdot^{\I_\B})$ as in the completeness proof. By Lemma~\ref{lem::TruthLemma}, $rep(a)\in C^{\I_\B}$, where $a$ is the root individual. The domain size is bounded by the number $m$ of individuals on $\B$, which by Lemma~\ref{lem::NumberofPrefixesInBranch} is exponential in the size of $C$ (or $C$ and $\mathcal{O}$).
\end{proof}




\section{Implementation and Experiments}\label{sec:implement}

In this section, we describe our implementation of the three tableau calculi and present their evaluation.
We have  implemented all three calculi in a single prover written in Python. 
Our implementation takes as input an \ALCi{}  concept, possibly together with an ABox and a TBox, however our preliminary experiments consider satisfiability of concepts without ontologies.
The prover parses the input concept using the Python library Lark, applies the tableau procedure, and outputs
information about the satisfiability of the concept.
The code and instructions for using the prover are available on https://github.com/ExtenDD/two-types-of-DDs-AAAI-2026.

The primary aim of our experiment is to  compare the efficiency of the prover depending on the amount of global descriptions (GDs) and local descriptions (LDs) in the input concepts.
All the experiments were run on a machine with the processor AMD Ryzen 5 PRO 7540U and 16GB RAM under Windows 11.

\paragraph{Generator} Our generator first builds a random binary syntax tree containing a predefined number of nodes, each corresponding to a subconcept; then, atomic concepts are randomly distributed among the leaves, binary operators (including GDs) among the inner nodes, and unary operators (including LDs) among all the nodes. 
The generator allows to customise the number of different atoms occurring in a concept, the number of GDs, LDs, and existential restrictions, as well as the chance for each subconcept to be preceded by  negation.

\paragraph{Experiments and results}
In all  experiments, the prover was run 5 times for each concept and the minimum was recorded as runtime, with the ``time-out'' limit set to 10s. 
The reported average runtimes do not consider the ``time-outs'', as well as concepts of the form $\neg \exists r.C$, which turn out satisfiable without applying any rule.
%
To compare runtimes for GDs and LDs we generated six datasets, each with 150 concepts containing a random number of atoms between 10 and 200.
First three datasets have concepts with GDs, but with no LDs.
The latter three datasest have concepts with  LDs, but with no GDs. 
For $k$ being the number of binary operators in a concept, the first three datasets contain $0.1\cdot k$, $0.3\cdot k$, and $0.5\cdot k$ GDs, respectively, whereas the next three datasets contain $0.1\cdot k$, $0.3\cdot k$, and $0.5\cdot k$ LDs.
Other parameters are kept the same in all datasets: the number of existential restrictions is $30\%$ of the number of subconcepts, 
and the number of different atoms is $50\%$ of the number of occurrences of atoms. 


\begin{table}[tbp]
    \small 
    \centering    
    {\setlength{\tabcolsep}{5pt}
    \begin{tabular}{llccc}
        \toprule
        \multicolumn{2}{c}{No. DDs in a concept:} & $0.1\cdot k$ & $0.3\cdot k$ & $0.5\cdot k$ \\
        \midrule
        \multirow{3}{*}{Global DDs} & \textbf{runtime avg.:} & 0.239s & 0.750s & 0.777s \\
        & runtime std.dev.: & 0.879s & 2.16s & 1.81s \\
        & no. of time-outs: & 21 & 31 & 42 \\
        \midrule
        \multirow{3}{*}{Local DDs} & \textbf{runtime avg.:} & 0.371s & 0.356s & 0.411s \\
        & runtime std.dev.: & 1.31s & 0.92s & 1.59s \\
        & no. of time-outs: & 15 & 32 & 28 \\
        \bottomrule
    \end{tabular}}
    \caption{Runtime for concepts with only GDs and only LDs; $k$ represents the number of binary operators in a concept}
    \label{tab1}
\end{table}

The runtimes for all six datasets are presented in  \Cref{tab1}. We do not include the time required for concept generation and for their parsing, as our focus was on the performance  of the implemented tableau procedure.\footnote{The runtime for parsing and concept generation grow linearly with concept size.  For example, concepts with 100 atoms require approx. 0.003s to be generated and approx. 0.5s to parse and concepts with 200 atoms require 0.006s and 1s, respectively.}
Two observations can be made: runtime and the number of time-outs are proportional to the number of descriptions in a concept (although this tendency is less visible for LDs), and greater for GDs than for LDs. 
The fact that GDs are more challenging aligns with our theoretical findings, whereas the fact that the growth of runtime is only linear, suggests practical feasibility of reasoning with DDs.

\begin{figure}[t]
\centering
\includegraphics[width = \linewidth]{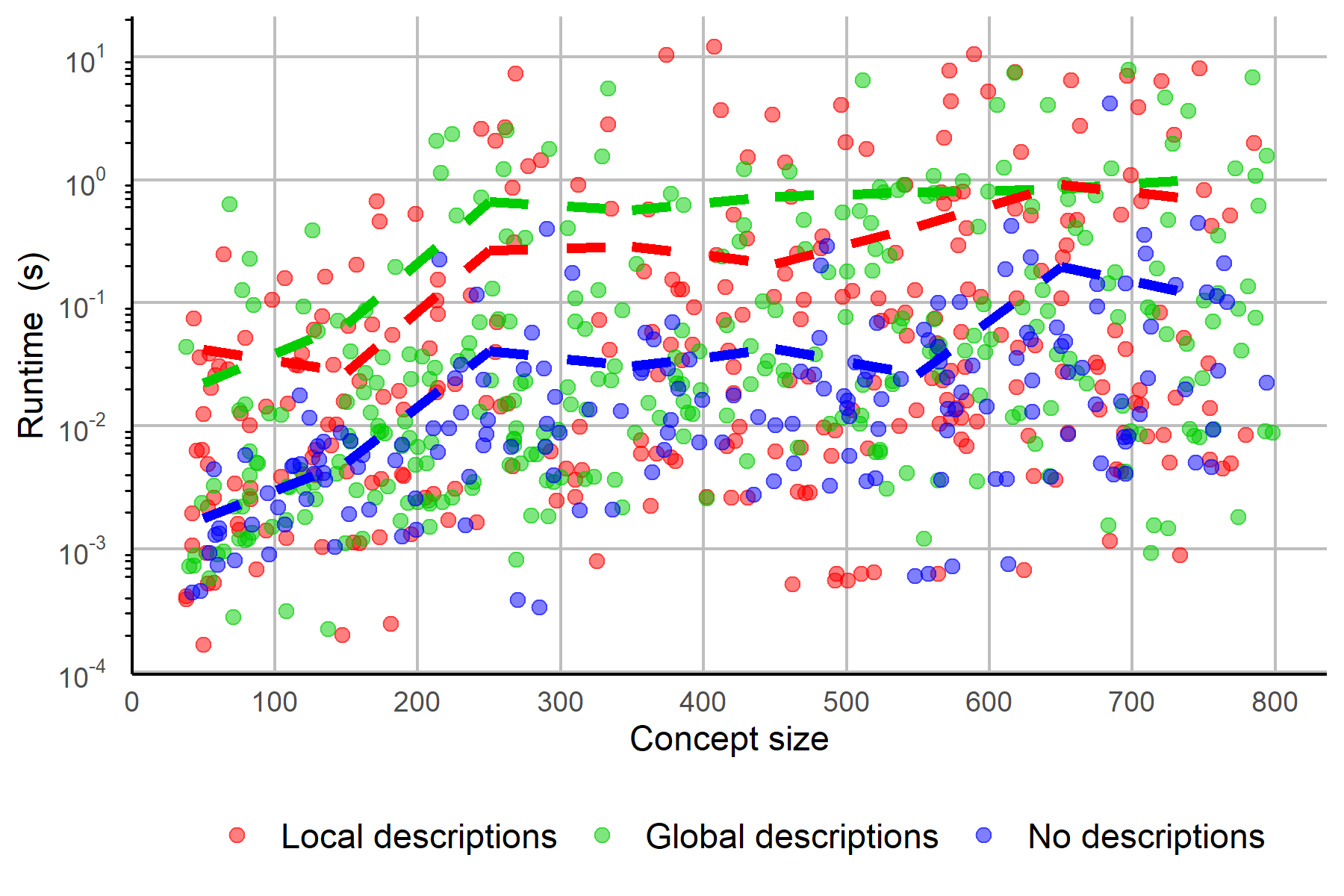}
\caption{Runtimes for concepts with only GDs, only LDs, and without DDs; dashed lines represent average runtimes}\label{runtime_vs_concept_size}
\end{figure}



In the next experiment we analysed the scalability of the prover, by testing its performance on concepts of growing size.
Recall that we identify the size of a concept with 
the number of symbols it uses, excluding parentheses.
For this, we grouped the previously generated concepts into a set with LDs only and a set with GDs only. 
Additionally,  we generated a set of 200 concepts without descriptions, leaving the other parameters the same as in the case of the other datasets. 
Results are depicted in \Cref{runtime_vs_concept_size}.
We observe that in all three cases, the runtime seems to grow polynomially with concept size, even though computational complexity of the satisfiability problem is \EXPT{}-complete. This may be due to how we generate concepts or to insufficient data.
Again, concepts with GDs  have higher runtimes than those with LDs, whereas  concepts without descriptions have clearly lower runtime, with only two time-outs across the whole dataset.

\section{Conclusions}

In this paper, we introduced three extensions of the standard description logic \ALC{}: \ALCil{} with local definite descriptions $\{\iota C\}$, \ALCig{} with global descriptions $\iota C.D$, and \ALCi{}, which supports both. We showed that all three logics are \EXPT{}-complete, but differ in expressive power: $\ALCil < \ALCig = \ALCi$. This expressiveness result is established via tailored bisimulations developed and analysed in the paper.
We also proposed tableau-based decision procedures for all the logics and implemented them. Experimental results show that definite descriptions increase reasoning time, with global descriptions incurring a higher cost than local ones. Nonetheless, the overhead remains manageable, confirming the practical feasibility of our extensions.
In future work, our aim is to optimise both the algorithms and their implementations.

\section*{Acknowledgements}

This research is funded by the European Union (ERC, ExtenDD, project number: 101054714). Views and opinions expressed are however those of the authors only and do not necessarily reflect those of the European Union or the European Research Council. Neither the European Union nor the granting authority can be held responsible for them.

\bibliography{DDbiblio}

@book{baader2017introduction,
  title={An introduction to description logic},
  author={Baader, Franz and Horrocks, Ian and Lutz, Carsten and Sattler, Uli},
  year={2017},
  publisher={Cambridge University Press}
}

@book{DBLP:books/daglib/0067423,
  author       = {Chen C. Chang and
                  H. Jerome Keisler},
  title        = {Model theory, Third Edition},
  series       = {Studies in logic and the foundations of mathematics},
  volume       = {73},
  publisher    = {North-Holland},
  year         = {1992}
}

@book{baader2003description,
  title={The description logic handbook: Theory, implementation and applications},
  author={Baader, Franz},
  year={2003},
address = {Cambridge},
  publisher={Cambridge University Press}
}

@article{russell1905denoting,
  title={On denoting},
  author={Russell, Bertrand},
  journal={Mind},
  volume={14},
  number={56},
  pages={479--493},
  year={1905},
  publisher={JSTOR}
}

@inproceedings{walega2023,
  title={Hybrid Modal Operators for Definite Descriptions},
  author={Wałęga, Przemysław Andrzej and Zawidzki, Michał},
  booktitle={Proc. of JELIA},
  pages={712--726},
  year={2023},
}

@inproceedings{borgida2016referring2,
  title={On referring expressions in information systems derived from conceptual modelling},
  author={Borgida, Alexander and Toman, David and Weddell, Grant},
  booktitle={Proc. of ER},
  pages={183--197},
  year={2016},
}

@inproceedings{toman2019identity,
  title={Identity resolution in ontology based data access to structured data sources},
  author={Toman, David and Weddell, Grant},
  booktitle={Proc. of PRICAI},
  pages={473--485},
  year={2019},
}

@inproceedings{toman2019finding,
  title={Finding ALL answers to {OBDA} queries using referring expressions},
  author={Toman, David and Weddell, Grant},
  booktitle={Proc. of AI},
  pages={117--129},
  year={2019},
}

@inproceedings{toman2018identity,
  title={Identity Resolution in Conjunctive Querying over {DL}-Based Knowledge Bases.},
  author={Toman, David and Weddell, Grant},
  booktitle={Proc. of DL},
  year={2018},
pages = {1--12}
}

@inproceedings{areces2008referring,
  title={Referring expressions as formulas of description logic},
  author={Areces, Carlos and Koller, Alexander and Striegnitz, Kristina},
  booktitle={Proc. of INLG},
  pages={42--49},
  year={2008}
}

@inproceedings{borgida2016referring,
  title={On referring expressions in query answering over first order knowledge bases},
  author={Borgida, Alexander and Toman, David and Weddell, Grant},
  booktitle={Proc. of KR},
pages = {319--328},
  year={2016}
}

@inproceedings{neuhaus2020free,
  title={Free Description Logic for Ontologists.},
  author={Neuhaus, Fabian and Kutz, Oliver and Righetti, Guendalina},
  booktitle={Proc. of JOWO},
  year={2020}
}

@inproceedings{artale2021free,
  title={On Free Description Logics with Definite Descriptions},
  author={Artale, Alessandro and Mazzullo, Andrea and Ozaki, Ana and Wolter, Frank},
  booktitle={Proc. of KR},
  pages={63--73},
  year={2021}
}

@book{blackburn2002modal,
  title={Modal Logic},
  author={Blackburn, Patrick and De Rijke, Maarten and Venema, Yde},
  year={2002},
publisher = {Cambridgde University Press},
address = {Cambridge},
series = {Cambridge Tracts in Theoretical Computer Science},
  volume = {53}
}

@inproceedings{Schild91,
  author       = {Klaus Schild},
  title        = {A Correspondence Theory for Terminological Logics: Preliminary Report},
  booktitle    = {Proc. of IJCAI},
  pages        = {466--471},
  year         = {1991},
  url          = {http://ijcai.org/Proceedings/91-1/Papers/072.pdf}
}

@incollection{Lambert2001,
	author	=	{Karel Lambert},
	booktitle = {New Essays in Free Logic},
	pages	=	{37--48},
	title	=	{Free Logic and Definite Descriptions},
	year	=	{2001},
	volume = {23},
}

@book{Rosser78,
author = {J. Barkley Rosser},
title = {Logic for Mathematicians},
year = {1978},
publisher = {Dover Publications},
address = {Dover}
}

@book{HilBer68,
author = {David Hilbert and Paul Bernays},
title = {Grundlagen der Mathematik I},
year = {1968},
publisher = {Springer},
address = {Berlin, Heidelberg}
}

@incollection{PelLin2005,
author = {Francis Jeffry Pelletier and Bernard Linsky},
title = {What is {F}rege's Theory of Descriptions},
booktitle = {On Denoting: 1905--2005},
year = {2005},
pages = {195--250}
}

@InProceedings{Bohrer2019,
author="Bohrer, Rose
and Fern{\'a}ndez, Manuel
and Platzer, Andr{\'e}",
editor="Fontaine, Pascal",
title="$\sf d{L}_\iota$: Definite {D}escriptions in {D}ifferential {D}ynamic {L}ogic",
booktitle="Proc. of CADE",
year="2019",
pages="94--110"
}

@article{Oppenheimer2011,
author = {Paul E. Oppenheimer and Edward N. Zalta and},
title = {A Computationally-Discovered Simplification of the Ontological Argument},
journal = {Australasian Journal of Philosophy},
volume = {89},
number = {2},
pages = {333--349},
year = {2011}
}

@article{Benzmuller2020,
title = {Automating Free Logic in {HOL}, with an Experimental Application in Category Theory},
journal = {Journal of Automated Reasoning},
author = {Benzm\"uller, Christoph and Scott, Dana S.},
volume = {64},
number = {1},
year = {2020},
pages = {53--72}
}

@inproceedings{Walega2024,
    title     = {{Expressive Power of Definite Descriptions in Modal Logics}},
    author    = {Wałęga, Przemysław Andrzej},
    booktitle = {Proc. of KR},
    pages     = {687--696},
    year      = {2024},
    month     = {8}
  }

@inproceedings{Indrzejczak2023a,
  author       = {Andrzej Indrzejczak},
  title        = {Towards Proof-Theoretic Formulation of the General Theory of Term-Forming
                  Operators},
  booktitle    = {Proc. of TABLEAUX},
  pages        = {131--149},
  year         = {2023}
}

@inproceedings{Indrzejczak2023b,
  author       = {Andrzej Indrzejczak and
                  Nils K{\"{u}}rbis},
   title        = {A Cut-Free, Sound and Complete Russellian Theory of Definite Descriptions},
  booktitle    = {Proc. of TABLEAUX},
  pages        = {112--130},
  year         = {2023}
}

@article{Indrzejczak2023c,
  author       = {Andrzej Indrzejczak},
  title        = {Russellian Definite Description Theory -- a Proof Theoretic Approach},
  journal      = {Review of Symbolic Logic},
  volume       = {16},
  number       = {2},
  pages        = {624--649},
  year         = {2023}
}

@inproceedings{Indrzejczak2024a,
  author       = {Andrzej Indrzejczak and
                  Yaroslav I. Petrukhin},
  title        = {Bisequent Calculi for Neutral Free Logic with Definite Descriptions},
  booktitle    = {Proc. of ARQNL},
  pages        = {48--61},
year = {2024}
}

@inproceedings{Indrzejczak2021a,
  author       = {Andrzej Indrzejczak and
                  Michal Zawidzki},
  title        = {Tableaux for Free Logics with Descriptions},
  booktitle    = {Proc. of TABLEAUX},
  pages        = {56--73},
  year         = {2021}
}

@article{Kurbis2025,
  author       = {Nils K{\"{u}}rbis},
  title        = {Normalisation for Negative Free Logics without and with Definite Descriptions},
  journal      = {Review of Symbolic Logic},
  volume       = {18},
  number       = {1},
  pages        = {240--272},
  year         = {2025}
}

@article{Kurbis2019a,
title={A Binary Quantifier for Definite Descriptions in Intuitionist Negative Free Logic: Natural Deduction and Normalisation},
volume={48},
number={2},
journal={Bulletin of the Section of Logic},
author={K\"urbis, Nils},
year={2019},
pages={81--97}
}

@article{Kurbis2019b,
title={Two Treatments of Definite Descriptions in Intuitionist Negative Free Logic},
volume={48},
number={4},
journal={Bulletin of the Section of Logic},
author={K\"urbis, Nils},
year={2019},
pages={299--317}
}

@article{Indrzejczak2023d,
author = {Indrzejczak, Andrzej and Zawidzki, Micha\l},
title = {When iota meets lambda},
journal = {Synthese},
volume = {201},
number = {1},
year = {2023},
note = {article number: 72}
}

@article{Indrzejczak2023e,
author = {Indrzejczak, Andrzej and Zawidzki, Micha\l},
title = {Definite descriptions and hybrid tense logic},
journal = {Synthese},
volume = {202},
number = {3},
year = {2023},
note = {article number: 98}
}

@article{Orlandelli2021,
    author = {Orlandelli, Eugenio},
    title = {Labelled calculi for quantified modal logics with definite descriptions},
    journal = {Journal of Logic and Computation},
    volume = {31},
    number = {3},
    pages = {923--946},
    year = {2021}
}

@book{Fitting2023,
author = {Melvin Fitting and Richard L. Mendelsohn},
title = {First-Order Modal Logic},
publisher = {Springer},
address = {Cham},
year = {2023},
edition = {2},
series = {Synthese Library},
volume = {480}
}

@article{Kaminski2009,
title = {Hybrid Tableaux for the Difference Modality},
journal = {Electronic Notes in Theoretical Computer Science},
volume = {231},
pages = {241--257},
year = {2009},
note = {Proc of. M4M5 2007},
author = {Mark Kaminski and Gert Smolka}
}

@inproceedings{Ren2010,
  author       = {Yuan Ren and
                  Kees van Deemter and
                  Jeff Z. Pan},
  title        = {Charting the Potential of Description Logic for the Generation of
                  Referring Expressions},
  booktitle    = {Proc. of INLG},
  year         = {2010},
pages = {115--123}
}

@inproceedings{Borgidaijcai2017,
  author    = {Alexander Borgida and David Toman and Grant Weddell},
  title     = {Concerning Referring Expressions in Query Answers},
  booktitle = {Proc. of IJCAI-17},
  pages     = {4791--4795},
  year      = {2017}
}

@inproceedings{TomanTreeModel16,
  author       = {David Toman and
                  Grant Weddell},
  title        = {Ontology Based Data Access with Referring Expressions for Logics with
                  the Tree Model Property -- (Extended Abstract)},
  booktitle    = {Proc. of {AI}},
  pages        = {353--361},
  year         = {2016}
}


\clearpage

\onecolumn
\appendix

\section*{Proof Details}

\section*{Proofs for \Cref{sec:exress}}

\noindent\textbf{Theorem~\ref{complexity}.}\quad
In all three logics $\ALCil$, $\ALCig$, and $\ALCi$,
both  satisfiability of concepts and satisfiability of ontologies is \EXPT{}-complete.
\medskip
\begin{proof}
For the upper bounds, we observe that
concept satisfiability is not harder than ontology satisfiability, and that 
$\ALCil$ and  $\ALCig$ are syntactical fragments of 
$\ALCi$.
Hence, to show all upper bounds it  suffices to show that satisfiability of $\ALCi$ ontologies is in \EXPT{}. To this end, we show a polynomial reduction to ontology satisfiability in $\ALCOiu$, which is  $\EXPT$-complete \cite[Thm. 2]{artale2021free}.
Local definite descriptions $\{\iota C\}$ are allowed in $\ALCOiu$, so it suffices to show how to translate global definite descriptions $\{\iota C.D\}$. 
Our reduction replaces each
$\iota C.D$ with $\exists u. ( \{ \iota C \} \sqcap D)$, where $u$ is the universal role.
The obtained ontology is  equivalent to the original ontology.
Moreover,  the translation is  feasible in polynomial
time because there are linearly many replacements; more precisely, the number of replacements is equal to the number of global definite descriptions in the input ontology and each replacement increases the number of symbols by 7.
Hence we obtain a polynomial reduction.


For the lower bounds, we observe that ontology satisfiability is not easier than concept satisfiability. Moreover, $\ALCil$ and  $\ALCig$ are syntactical fragments of 
$\ALCi$. Hence, it suffices to show \EXPT{}-hardness of concept satisfiability in  $\ALCil$ and  $\ALCig$. 
For this, we show log-space reductions of the consistency checking of an $\ALC$  concept $C$ with respect to an $\ALC$ TBox $\T$, which is \EXPT{}-complete 
\cite[Thm. 3.27]{baader2003description}.
In the case of \ALCil{}, we construct a concept $C'$ as follows:
$$C'  \coloneqq 
C \sqcap 
\! \bigsqcap_{(D \sqsubseteq E) \in \T} \!
\Big( 
(\neg D \sqcup E) \sqcap 
\{ \iota  (\neg (\neg D \sqcup E)  \sqcup A_{D \sqsubseteq E} ) \}  
\Big)
,
$$
where $A_{D \sqsubseteq E}$ is a fresh atomic concept for each axiom $D \sqsubseteq E$ in $\T$.
Hence $\I \models C'(d)$ if $d$ satisfies $C$ and all axioms in $\T$.
Moreover,  satisfaction of  definite descriptions  implies  that  $\I \models A_{D \sqsubseteq E}(d)$, so axioms $D \sqsubseteq E$ need to hold in all other individuals than $d$. As a result, $C'$ is satisfiable if and only if $C$ is satisfiable with respect to $\T$.
In the case of $\ALCig$, we construct a concept $C''$ as follows:
$$C'  \coloneqq 
C \sqcap 
\! \bigsqcap_{(D \sqsubseteq E) \in \T} \!
\Big( 
(\neg D \sqcup E)  \sqcap
A_{D \sqsubseteq E} \sqcap
\{ \iota  (\neg (\neg D \sqcup E)  \sqcup A_{D \sqsubseteq E} ) . \top \}  
\Big)
,
$$
which allows us to obtain a similar behaviour as in the case of $\ALCi$.
In particular, 
$\I \models C''(d)$ if $d$ satisfies $C$, all axioms in $\T$, and all $A_{D \sqsubseteq E}$.
Then, by the  satisfaction of  definite descriptions at $d$, we obtain that all  axioms $D \sqsubseteq E$ need to hold in all other individuals than $d$.
Hence
$C''$ is satisfiable if and only if $C$ is satisfiable with respect to $\T$.
\end{proof}

\noindent\textbf{Proposition~\ref{translations}.}\quad
There is an exponential equivalence-preserving translation of $\ALCil$ concepts into $\ALCig$ concepts.
There is also a polynomial translation of 
$\ALCil$ ontologies into conservative extensions in $\ALCig$.
\medskip
\begin{proof}[Proof sketch]
The exponential translation is obtained by simply  replacing each
$\{ \iota C \}$, such that $C$ does not mention definite descriptions, with $C \sqcap \iota C. \top$.
This clearly an equivalence preserving translation.
Although there are linearly many replacements (as many as definite descriptions in the input concept), translation of $\{ \iota C \}$ leads to two new occurrences of $C$, as in $C \sqcap \iota C. \top$.
As a result, the translation is exponential.

The polynomial translation replaces each $\{ \iota C \}$, where $C$ is a concept not mentioning definite descriptions, with $A_C \sqcap \iota A_C. \top$, where $A_C$ is a fresh concept name, and adds axioms $A_C \equiv C$.
We observe that the obtained ontology is a conservative extension of the input one. Moreover, the translation is polynomial, because it requires linearly many replacements. Note that similarly as in the exponential translation
each replacement leads to two occurrences of $A_C$ in the concept. However, now those are atomic concepts which does not lead to exponential increase of the concept.
\end{proof}

\noindent\textbf{Claim A.}\quad
Let $\ST_x$ be a translation from $\ALCii$ to $\FOtwo$, defined as follows, where $\ST_y$ is an analogous translation but it uses $y$ in the place of $x$ (and $x$ in the place of $y$):
\[
\begin{array}{rcl}
\ST_x(A) &=& A(x), \\
\ST_x(\neg C) &=& \neg\,\ST_x(C), \\
\ST_x(C \sqcap D) &=& \ST_x(C) \land \ST_x(D), \\
\ST_x(\exists r . C) &=& \exists y\,\big( r(x,y) \land \ST_y(C)\big), \\
\ST_x(\{\iota C\}) &=& \ST_x(C) \land \forall y\big(\ST_y(C) \rightarrow x = y\big), \\
\ST_x(\iota C . D) &=& \exists y\,\big( \ST_y(C) \land \forall x\big(\ST_x(C) \rightarrow x = y\big) \land \ST_y(D)\big).
\end{array}
\]
Then, for any $\ALCii$-interpretation $\I = (\Delta^\I, \cdot^\I)$, any $d \in \Delta^\I$, and any concept $C$, we have:
\[
d \in C^\I \quad\text{iff}\quad \I \models \ST_x(C)[x \mapsto d].
\]

\begin{proof}
    We conduct the proof by structural induction on the concept $C$.\smallskip

    \noindent \textbf{Case} $C=A$
    \begin{align*}
    d\in A^\I&\text{ iff }\I\models A(d)\I\\
    &\text{ iff }\I\models A(x)[x\mapsto d]\\
    &\text{ iff }\I\models\ST_x(C)[x\mapsto d].
    \end{align*}

    \noindent \textbf{Case} $C=\neg D$
    \begin{align*}
        d\in(\neg D)^\I&\text{ iff }d\notin D^\I\\
        &\text{ iff }\I\not\models\ST_x(D)[x\mapsto d]\tag{ind. hyp.}\\
        &\text{ iff }\I\models\neg\ST_x(D)[x\mapsto d]\\
        &\text{ iff }\I\models\ST_x(\neg D)[x\mapsto d].
    \end{align*}

    \noindent \textbf{Case} $C=D\sqcap E$
    \begin{align*}
        d\in(D\sqcap E)^\I&\text{ iff }d\in D^\I\text{ and }d\in E^\I\\
        &\text{ iff }\I\models\ST_x(D)[x\mapsto d]\text{ and }\I\models\ST_x(E)[x\mapsto d]\tag{ind. hyp.}\\
        &\text{ iff }\I\models\neg\ST_x(D)[x\mapsto d]\land\ST_x(E)[x\mapsto d]\\
        &\text{ iff }\I\models\ST_x(D\land E)[x\mapsto d].
    \end{align*}

    \noindent \textbf{Case} $C=\exists r.D$
    \begin{align*}
        d\in(\exists r.D)^\I&\text{ iff }\text{for some }e\in\Delta^\I\text{ such that } r(d,e), e\in D^\I\\
        &\text{ iff }\I\models r(x,y)[x\mapsto d,y\mapsto e]\text{ and }\I\models\ST_y(D)[y\mapsto e]\\
        &\text{ iff }\I\models (r(x,y)\land\ST_y(D))[x\mapsto d,y\mapsto e]\tag{ind. hyp.}\\
        &\text{ iff }\I\models\exists y(r(x,y)\land\ST_y(D))[x\mapsto d]\\
        &\text{ iff }\I\models\ST_x(\exists r.D)[x\mapsto d].
    \end{align*}

    \noindent \textbf{Case} $C=\{\iota D\}$
    \begin{align*}
        d\in(\{\iota D\})^\I&\text{ iff }\{d\}= D^\I\\
        &\text{ iff }d\in D^\I\text{ and for any }e\in\Delta^\I,\text{ if }e\in D^\I,\text{ then }e=d\\
        &\text{ iff }\I\models \ST_x(D)[x\mapsto d]\text{ and }\forall y(\ST_y(D)\to y=x)[x\mapsto d]\tag{ind. hyp.}\\
        &\text{ iff }\I\models(\ST_x(D)\land\forall y(\ST_y(D)\to y=x))[x\mapsto d]\\
        &\text{ iff }\I\models\ST_x(\{\iota D\})[x\mapsto d].
    \end{align*}

    \noindent \textbf{Case} $C=\iota D.E$
    \begin{align*}
        d\in(\iota D.E)^\I&\text{ iff }\text{ for some }e\in\Delta^\I,\{e\}= D^\I\text{ and }e\in E^\I\\
        &\text{ iff }\text{ for some }e\in\Delta^\I,e\in D^\I\text{ and for any }e'\in\Delta^\I,\text{ if }e'\in D^\I,\text{ then }e'=e,\text{ and }e\in E^\I\\
        &\text{ iff }\I\models \ST_y(D)[y\mapsto e]\text{ and }\I\models\forall x(\ST_x(D)\to x=y)[y\mapsto e]\text{ and }\I\models\ST_y(E)[y\mapsto e]\tag{ind. hyp.}\\
        &\text{ iff }\I\models (\ST_y(D)\land\forall x(\ST_x(D)\to x=y)\land\ST_y(E))[y\mapsto e]\\
        &\text{ iff }\I\models \exists y(\ST_y(D)\land\forall x(\ST_x(D)\to x=y)\land\ST_y(E))\\
        &\text{ iff }\I\models \exists y(\ST_y(D)\land\forall x(\ST_x(D)\to x=y)\land\ST_y(E))[x\mapsto d]\tag{$x$ not free}\\
        &\text{ iff }\I\models \ST_x(\iota D.E)[x\mapsto d]
    \end{align*}
    This completes the proof.
\end{proof}

\noindent\textbf{Theorem~\ref{bisimulation_thm}.}\quad
For all pointed  interpretations $(\mathcal{I}, d)$ and $(\mathcal{J}, e)$, and both $\Lan \in \{ \ALCil, \ALCig \}$ the following hold:
\begin{enumerate}
\item[1.] if $(\I, d) \sim_{\Lan} (\J, e)$, then $(\I, d) \equiv_{\Lan} (\J, e)$,
\item[2.] if $(\I, d) \equiv_{\Lan} (\J, e)$ and $\I, \J$ are $\omega$-saturated, then $(\I, d) \sim_{\Lan} (\J, e)$.
\end{enumerate}
\medskip
\begin{proof}
First, we show Statement 1. Assume that $(\I,d)$ and $(\J,e)$ are pointed interpretations and there exists an $\Lan$-bisimulation $Z$ such that $(d,e)\in Z$. We need to show that for any $\Lan$-concept $C$, $d\in C^\I$ iff $e\in C^\J$. 
We simultaneously prove the statement for both $\Lan \in \{ \ALCil, \ALCig \}$ by induction on the structure of $C$, adjusting the type of $Z$ accordingly.\smallskip

\noindent
\textbf{Case} $C=A$\quad Since $(\I,d)\sim^\Lan(\J,e)$, by condition \textbf{Atom}, $d\in A^\I$ iff $e\in A^\J$.\smallskip

\noindent\textbf{Case} $C=\neg D$\quad Let $d\in(\neg D)^\I$. By the semantic condition for $\neg$ it means that $d\notin D^\I$. By the induction hypothesis we obtain $e\notin D^\J$, and consequently, $e\in(\neg D)^\J$, as expected. The opposite implication can be proved analogously.\smallskip

\noindent
\textbf{Case} $C= D\sqcap E$\quad  Let $d\in(D\sqcap E)^\I$. By the semantic condition for $\sqcap$, $d\in D^\I$ and $d\in E^\I$. By the induction hypothesis, $e\in D^\J$ and $e\in E^\J$, from which $e\in (D\sqcap E)^\J$ follows. The converse implication can be proved similarly.\smallskip

\noindent
\textbf{Case} $C=\exists r.D$\quad Let $d\in(\exists r.D)^\I$. This means that there exists an individual $d'\in\Delta^\I$ such that $(d,d')\in r$ and $d'\in D^\I$. By condition \textbf{Forth} there must exist $e'\in\Delta^\J$ such that $(e,e')\in r$ and $(d',e')\in Z$. We can then apply the induction hypothesis to obtain $e'\in D^\J$. By the semantic condition for $\exists r$ we get $e\in(\exists r.D)^\J$, as required. The converse implication can be proved similarly, but with condition \textbf{Back} used instead of condition \textbf{Forth}.\smallskip

The cases specific to $\ALCil$ and $\ALCig$ are as follows:
\smallskip

\noindent
\textbf{Case} $C = \{\iota D\}$\quad 
Let $d \in (\{\iota D\})^\I$. By the semantics of DDs, we have $d \in D^\I$ and $|D^\I| = 1$. By the induction hypothesis, $e \in D^\J$. Since $d \in \Dom(Z)$ and $D \in \names(\Dom(Z), \I)$, condition \textbf{Names}$_L$ implies $D \in \names(\Rng(Z), \J)$. Hence, there exists an individual $e' \in \Rng(Z)$ such that $D^\J = \{e'\}$. Since $e \in \Rng(Z)$, it follows that $e = e'$, and thus $e \in (\{\iota D\})^\J$.
Similarly $e \in (\{\iota D\})^\J$ implies $d \in (\{\iota D\})^\I$.
\smallskip

\noindent
\textbf{Case} $C = \iota D.E$\quad 
Let $d \in (\iota D.E)^\I$. Then $|D^\I| = 1$ and $D^\I \subseteq E^\I$. This implies $|(D \sqcap E)^\I| = 1$, so both $D$ and $D \sqcap E$ are in $\names(\Delta^\I, \I)$. By condition \textbf{Names}$_G$, we obtain $D, D \sqcap E \in \names(\Delta^\J, \J)$, whence $|D^\J| = 1$ and $D^\J \subseteq E^\J$. Therefore, $e \in (\iota D.E)^\J$ by the semantics of $\iota$.
Similarly $e \in (\{\iota D.E\})^\J$ implies $d \in (\{\iota D.E\})^\I$.
\smallskip

Now we prove Statement~2 for $\Lan = \ALCig$.

Assume $(\I, d) \equiv_{\ALCig} (\J, e)$ and that $\I, \J$ are $\omega$-saturated. 
Since $(\I, d) \equiv_{\ALCig} (\J, e)$ implies $(\I, d) \equiv_{\ALC} (\J, e)$, by the standard results on \ALC{} bisimulations~\cite[Thm.~2.20]{blackburn2002modal}, \cite[Thm.~1]{Schild91}, there exists an \ALC{} bisimulation $Z$ between $\I$ and $\J$ with $(d, e) \in Z$.

We claim that $Z$ is also an \ALCig{} bisimulation, i.e.,
\[
\names(\Delta^\I, \I) = \names(\Delta^\J, \J).
\]
Indeed, suppose $C \in \names(\Delta^\I, \I)$, so $d \in (\iota C . \top)^\I$. 
Since $(\I, d) \equiv_{\ALCig} (\J, e)$, it follows that $e \in (\iota C . \top)^\J$, hence $C \in \names(\Delta^\J, \J)$. 
The converse inclusion is symmetric, so the sets of names coincide.

For $\Lan = \ALCil$, define
\[
Z = \{(k, l) \in \Delta^\I \times \Delta^\J \mid (\I, k) \equiv_{\ALCil} (\J, l)\}.
\]
Clearly, $(d, e) \in Z$. We now show that $Z$ is an $\ALCil$-bisimulation.

\medskip
\noindent\textbf{Atom.} 
This is immediate: for every $(d', e') \in Z$ and every $\ALCil$-concept $C$, we have
\[
d' \in C^\I \quad\text{iff}\quad e' \in C^\J.
\]

\medskip
\noindent\textbf{Zig.} 
Let $(d', e') \in Z$ and suppose $(d', d'') \in r^\I$.  
Define the set of all concepts satisfied by $d''$ as
\[
\Sigma_{d''} = \{C \mid d'' \in C^\I \}.
\]
By the semantics of $\exists r$, we have $d' \in (\exists r . C)^\I$ for each $C \in \Sigma_{d''}$.  
Now consider the set of corresponding $\FOtwo$-formulas
\[
\Gamma_{d'} = \{\ST_x(\exists r . C)[x \mapsto d'] \;\big|\; C \in \Sigma_{d''}\}.
\]
Each $\varphi\in\Gamma_{d'}$ has the form $\exists y\,(r(d',y) \land \ST_y(C))$ for some $C \in \Sigma_{d''}$, meaning that the set of formulas
\[
\Pi_{d'}=\{r(d',y)\}\cup\{\ST_y(C)\mid C\in\Sigma_{d''}\}
\]
is realisable in $\I$.  
By the fact that $(\I,d')\equiv_{\ALCil}(\J,e')$ and by Claim A we infer that for each finite subset $\Pi_{d'}'\subseteq\Pi_{d'}$
\[
\Pi_{d'}'[e'\mapsto d']
\]
is realisable in $\J$. By the $\omega$-saturation of $\J$, it follows that $\Pi_{d'}[e'\mapsto d']$ is realisable in $\J$. Hence, there exists some $e'' \in \Delta^\J$ such that
\[
\J \models \{r(e',e'')\}\cup\{\ST_y(C)[y \mapsto e''] \mid C \in \Sigma_{d''}\}.
\]
By Claim A, this implies $e'' \in C^\J$ for all $C \in \Sigma_{d''}$. 
Therefore, by the definition of $Z$, we obtain $(d'', e'') \in Z$ as required.

\medskip
\noindent\textbf{Zag.} 
This direction is symmetric and follows by the same reasoning.

\medskip
\noindent\textbf{Names$_L$.} 
Assume $C \in \names(\Dom(Z), \I)$. Then there exists $d' \in \Dom(Z)$ such that 
\[
d' \in (\{\iota C\})^\I.
\]
Since $d' \in \Dom(Z)$, there exists $e' \in \Delta^\J$ with $(d', e') \in Z$.  
From $d' \in (\{\iota C\})^\I$ and $(d', e') \in Z$, it follows that $e' \in (\{\iota C\})^\J$, hence $C \in \names(\Rng(Z), \J)$.

Thus,
\[
\names(\Dom(Z), \I) \subseteq \names(\Rng(Z), \J).
\]
The reverse inclusion is analogous, and therefore
\[
\names(\Dom(Z), \I) = \names(\Rng(Z), \J).\qedhere
\]
\end{proof}

\noindent\textbf{Theorem~\ref{thm:total}.}\quad
Let $\I$ and $\J$ be  finite interpretations and let $\Delta' \subseteq \Delta^\I$ and $\Delta'' \subseteq \Delta^\J$.
If $\names(\Delta',\I) \neq \emptyset \neq \names(\Delta'',\J)$, then $\names(\Delta',\I) = \names(\Delta'',\J)$ if and only if 
the maximal \ALC{}-bisimulation $Z$ between $\I$ and $\J$ is a total relation and the restriction of $Z$ to $\named(\Delta',\I) \times \named(\Delta'',\J)$ is also total.
\medskip
\begin{proof}
We prove both directions of the equivalence.

\medskip
\noindent
\textbf{($\Rightarrow$)} Assume $\names(\Delta', \I) = \names(\Delta'', \J)$, and let $Z$ be the maximal \ALC-bisimulation between $\I$ and $\J$. We show:\vspace{.5em}

\noindent 1. $Z$ is total on $\Delta^\I \times \Delta^\J$, and\vspace{.5em}

\noindent 2. $Z$ is total on $\named(\Delta', \I) \times \named(\Delta'', \J)$.\vspace{.5em}

\noindent
\textit{(2) The restriction of $Z$ is total.}  
We first show left-totality; the right-totality is analogous.
Suppose for contradiction that there exists $d \in \named(\Delta', \I)$ with no $e \in \named(\Delta'', \J)$ such that $(d, e) \in Z$. 
Then there exists $C \in \names(\Delta', \I)$ such that $\{d\} = C^\I$. Since $\names(\Delta', \I) = \names(\Delta'', \J)$, there exists $e \in \named(\Delta'',\J)$ with $\{e\} = C^\J$. But $(d, e) \notin Z$, so by the Hennessy-Milner theorem for \ALC{}, 
there exists an \ALC{}-concept $D$ such that $d \in D^\I$ and $e \notin D^\J$.
Then, $\{d\} = (C \sqcap D)^\I$, so $C \sqcap D \in \names(\Delta', \I)$, while $(C \sqcap D)^\J = \emptyset$, contradicting the assumption that $\names(\Delta', \I) = \names(\Delta'', \J)$. Hence, such $e\in\named(\Delta'',\J)$, for which $(d,e)\in Z$ holds, must exist.\vspace{.5em}

\noindent
\textit{(1) $Z$ is total.}  
Let $d \in \Delta^\I \setminus \named(\Delta', \I)$ and suppose for contradiction that for all $e \in \Delta^\J$, $(d, e) \notin Z$.
Since $\names(\Delta'', \J) \neq \emptyset$, there exists a concept $C \in \names(\Delta'', \J) $ and an individual $e^* \in \named(\Delta'', \J)$ such that $\{e^*\} = C^\J$. For each $e \in \Delta^\J \setminus \{e^*\}$, the Hennessy-Milner theorem, together with the fact that $(d,e)\notin Z$, gives a concept $C_e$ such that $d \notin (C_e)^\I$ and $e \in (C_e)^\J$.
Define:
$D := C \sqcup \bigsqcap_{e \in \Delta^\J \setminus \{e^*\}} \neg C_e.$
Then $e^* \in D^\J$, and for all $e\in\Delta^\J\setminus\{e^*\}$, $e \notin D^\J$, so $\{e^*\} = D^\J$, implying $D \in \names(\Delta'', \J) = \names(\Delta', \I)$. Since $d \in D^\I$, we get $d \in \named(\Delta', \I)$, a contradiction.\vspace{.5em}

\noindent
\textbf{($\Leftarrow$)} Assume $Z$ is total on $\Delta^\I \times \Delta^\J$ and on $\named(\Delta', \I) \times \named(\Delta'', \J)$. We show $\names(\Delta', \I) = \names(\Delta'', \J)$.
Let $C \in \names(\Delta', \I)$, so there is $d \in \named(\Delta', \I)$ with $\{d\} = C^\I$. By totality, there is $e \in \named(\Delta'', \J)$ with $(d, e) \in Z$. Then $e \in C^\J$ by bisimulation invariance.
If $\{e\} = C^\J$, then $C \in \names(\Delta'', \J)$ as required. Suppose not; then there exists $e' \neq e$ with $e' \in C^\J$. By the totality of $Z$, there exists $d' \in \Delta^\I$ with $(d', e') \in Z$. We have two cases:\vspace{.5em}

\noindent 1. If $d' = d$: then for every concept $D$ such that $e \in D^\J$, we get $d \in D^\I$ and further $e' \in D^\J$ by bisimulation invariance, so there exists no concept $D$ such that $\{e\}=D^\J$, contradicting $e\in\named(\Delta'', \J)$.\vspace{.5em}

\noindent 2. If $d' \neq d$: then both $d, d' \in C^\I$ by bisimulation invariance, so $|C^\I| \geq 2$, contradicting $C \in \names(\Delta', \I)$.\vspace{.5em}

\noindent Thus, $C \in \names(\Delta'', \J)$, so $\names(\Delta', \I) \subseteq \names(\Delta'', \J)$. The reverse inclusion is shown similarly, hence $\names(\Delta', \I) = \names(\Delta'', \J)$.
\end{proof}

\newpage
\section*{Proofs for \Cref{sec:tableau}}

\noindent\textbf{Lemma~\ref{lem::UnsatisfiabilityOfU}.}\quad
    Let $(\mathsf{r})$ be a rule of $\TALCi$ applied to a branch $\B$, and let $\B_1,\ldots,\B_n \supseteq \B$ be the resulting branches. If $\B$ is satisfiable, then so is some $\B_i$ for $i \in \{1,\ldots,n\}$.

\begin{proof}
We consider each rule, omitting the ABox, clash, and propositional cases as straightforward.

\smallskip
\noindent\textbf{Case $(TBox)$} Let $\mathcal{P}r = \{C \sqsubseteq D \in TBox,\, a:E\}$ be satisfiable. Then there exists an interpretation $\I = (\Delta^\I, \cdot^\I)$ such that $C^\I \subseteq D^\I$ and $a \in E^\I$. We have two cases:
\begin{itemize}
    \item[(i)] If $a \in C^\I$, then $a \in D^\I$, so $a \notin (C \sqcap \neg D)^\I$.
    \item[(ii)] If $a \notin C^\I$, then $a \notin (C \sqcap \neg D)^\I$.
\end{itemize}
In both cases, $a \in \neg(C \sqcap \neg D)^\I$, hence $\mathcal{P}r \cup \{a:\neg(C \sqcap \neg D)\}$ is satisfiable.
\smallskip

\noindent\textbf{Case $(\neg\exists r)$} Let $\mathcal{P}r = \{a:\neg\exists r.C,\, r(a,a')\}$ be satisfiable. Then $a \in (\neg\exists r.C)^\I$ and $(a,a') \in r^\I$ imply $a' \in (\neg C)^\I$, so $\mathcal{P}r \cup \{a':\neg C\}$ is satisfiable.\smallskip

\noindent\textbf{Case $(\exists r)$} Let $\mathcal{P}r = \{a:\exists r.C\}$ be satisfiable. Then $a \in (\exists r.C)^\I$, so there exists $b \in \Delta^\I$ such that $(a,b) \in r^\I$ and $b \in C^\I$. Hence, $\mathcal{P}r \cup \{b:C,\, r(a,b)\}$ is satisfiable.\smallskip

\noindent\textbf{Case $(\iota_1^g)$} Let $\mathcal{P}r = \{a:\iota C.D\}$ be satisfiable. Then there exists $b \in \Delta^\I$ with $b \in C^\I \cap D^\I$, so $\mathcal{P}r \cup \{b:C,\, b:D\}$ is satisfiable.\smallskip

\noindent\textbf{Case $(\iota_2^g)$} Let $\mathcal{P}r = \{a:\iota C.D,\, a':C,\, a'':C,\, a':E\}$ be satisfiable. Since $a \in (\iota C.D)^\I$, there exists a unique $c \in \Delta^\I$ with $c \in C^\I$. As both $a'$ and $a''$ are in $C^\I$, we must have $a' = a'' = c$. Hence $a'' \in E^\I$ and $\mathcal{P}r \cup \{a'':E\}$ is satisfiable.\smallskip

\noindent\textbf{Case $(\neg\iota^g)$} Let $\mathcal{P}r = \{a:\neg\iota C.D,\, a':E\}$ be satisfiable. Then $a \in (\neg\iota C.D)^\I$ and $a' \in E^\I$, meaning no unique individual satisfies both $C$ and $D$. Three cases:
\begin{itemize}
    \item[(i)] $a' \notin C^\I$, so $a' \in (\neg C)^\I$ and $\mathcal{P}r \cup \{a':\neg C\}$ is satisfiable.
    \item[(ii)] $a' \notin D^\I$, so $a' \in (\neg D)^\I$ and $\mathcal{P}r \cup \{a':\neg D\}$ is satisfiable.
    \item[(iii)] $a' \in C^\I \cap D^\I$, and there exists $b' \in \Delta^\I$, $b' \neq a'$, such that $b' \in C^\I$. Extend $\I$ to $\I'$ so that $(A_C^g)^{\I'} = \{a'\}$. Then:
    \[
    a' \in (A_C^g)^{\I'}, \quad b' \in (\neg A_C^g)^{\I'},
    \]
    hence $\mathcal{P}r \cup \{b:C,\, b:A_C^g,\, b':C,\, b':\neg A_C^g\}$ is satisfiable.
\end{itemize}\smallskip

\noindent\textbf{Case $(cut_\iota^g)$} Let $\mathcal{P}r = \{a:\iota C.D,\, a':E\}$ be satisfiable. Then either $a' \in C^\I$ or $a' \in (\neg C)^\I$, so $\mathcal{P}r \cup \{a':C\}$ or $\mathcal{P}r \cup \{a':\neg C\}$ is satisfiable.\smallskip

\noindent\textbf{Case $(\iota_1^\ell)$} Let $\mathcal{P}r = \{a:\{\iota C\}\}$ be satisfiable. Then $\{a\} = C^\I$, so $a \in C^\I$ and $\mathcal{P}r \cup \{a:C\}$ is satisfiable.\smallskip

\noindent\textbf{Case $(\iota_2^\ell)$} Let $\mathcal{P}r = \{a:\{\iota C\},\, a':C,\, a'':C,\, a':E\}$ be satisfiable. Since $C^\I = \{a\}$, we have $a = a' = a''$, so $a'' \in E^\I$ and $\mathcal{P}r \cup \{a'':E\}$ is satisfiable.\smallskip

\noindent\textbf{Case $(\neg\iota^\ell)$} Let $\mathcal{P}r = \{a:\neg\{\iota C\}\}$ be satisfiable. Then $\{a\} \neq C^\I$, so:
\begin{itemize}
    \item[(i)] If $a \notin C^\I$, then $a \in (\neg C)^\I$, so $\mathcal{P}r \cup \{a:\neg C\}$ is satisfiable.
    \item[(ii)] Otherwise, there exists $b \neq a$ with $b \in C^\I$. Extend $\I$ to $\I'$ with $(A_C)^{\I'} = \{b\}$. Then:
    \[
    a \in (\neg A_C)^{\I'}, \quad b \in A_C^{\I'},
    \]
    hence $\mathcal{P}r \cup \{a:\neg A_C,\, b:C,\, b:A_C\}$ is satisfiable.
\end{itemize}\smallskip

\noindent\textbf{Case $(cut_\iota^\ell)$} Let $\mathcal{P}r = \{a:\{\iota C\},\, a':D\}$ be satisfiable. Then either $a' \in C^\I$ or $a' \in (\neg C)^\I$, so one of $\mathcal{P}r \cup \{a':C\}$ or $\mathcal{P}r \cup \{a':\neg C\}$ is satisfiable.
\end{proof}

\noindent\textbf{Claim B.}\quad
For any individual $a$ on $\B$, we have $\theory{\B}{a} = \theory{\B}{rep(a)}$.

\begin{proof}
If $rep(a) = a$, the claim is immediate. Otherwise, for some $C \in \mathsf{DD}(\B)$, both $a : C$ and $rep(a) : C$ are in $\B$. Then exhaustive application of $(\iota_2^g)$ or $(\iota_2^\ell)$ yields $\theory{\B}{a} = \theory{\B}{rep(a)}$.
\end{proof}

\noindent\textbf{Lemma~\ref{lem::TruthLemma}.}\quad
For any assertion $a : C \in \B$, $rep(a) \in C^{\I_\B}$.

\begin{proof}
By structural induction on $C$.\smallskip

\noindent\textbf{Case} $C = A$\quad Immediate by the definition of $\cdot^{\I_\B}$.\smallskip

\noindent\textbf{Case} $C = \neg A$\quad Toward a contradiction, suppose $a : \neg A \in \B$ but $rep(a) \notin (\neg A)^{\I_\B}$. Then $rep(a) \in A^{\I_\B}$, which, by the definition of $\cdot^{\I_\B}$, implies $rep(a): A \in \B$. By Claim B, it follows that $a : A \in \B$, contradicting the openness of $\B$.\smallskip

\noindent\textbf{Case} $C = D \sqcap E$\quad Assume $a : D \sqcap E \in \B$. Since $\B$ is saturated, rule $(\sqcap)$ must have been applied, yielding $a : D, a : E \in \B$. By the induction hypothesis, $rep(a) \in D^{\I_\B}$ and $rep(a) \in E^{\I_\B}$, hence $rep(a) \in (D \sqcap E)^{\I_\B}$.\smallskip

\noindent\textbf{Case} $C = \neg(D \sqcap E)$\quad Assume $a : \neg(D \sqcap E) \in \B$. As $\B$ is saturated, rule $(\neg\sqcap)$ must have been applied, yielding either $a : \neg D \in \B$ or $a : \neg E \in \B$. In the first case, by the induction hypothesis, $rep(a) \in (\neg D)^{\I_\B}$ and thus $rep(a) \in (\neg(D \sqcap E))^{\I_\B}$. The second case is analogous.\smallskip

\noindent\textbf{Case $C = \exists r . D$}\quad Assume $a : \exists r . D \in \B$. Since $\B$ is saturated, two cases arise:
\begin{itemize}
\item[(i)] If $(\exists r)$ was applied, then $\B$ contains $b : D$ and $r : (a, b)$. Thus $(rep(a), rep(b)) \in r^{\I_\B}$, and by the induction hypothesis, $rep(b) \in D^{\I_\B}$, hence $rep(a) \in (\exists r . D)^{\I_\B}$.
\item[(ii)] If $(\exists r)$ was blocked, then ($block_\exists$) must be satisfied. Then there exists an $(r,D)$-proxy successor of $a$ on $\B$, say $a'$. By definition, $a':D\in \B$. Thus $(rep(a), rep(a')) \in r^{\I_\B}$ and by the induction hypothesis, $rep(a') \in D^{\I_\B}$, so $rep(a) \in (\exists r . D)^{\I_\B}$.
\end{itemize}\smallskip

\noindent\textbf{Case} $C = \neg \exists r . D$\quad Assume $a : \neg \exists r . D \in \B$. We must show that for every $a'$ such that $(rep(a), rep(a')) \in r^{\I_\B}$, it holds that $rep(a') \in (\neg D)^{\I_\B}$. Let such an $a'$ be given. We distinguish two cases:
\begin{itemize}
\item[(i)] There exist $b, b'$ on $\B$ such that $b = rep(a)$, $b' = rep(a')$, and $r : (b, b') \in \B$. By Claim B, $\theory{\B}{b} = \theory{\B}{a}$, so $b : \neg \exists r . D \in \B$. Since $\B$ is saturated, rule $(\neg \exists r)$ must have been applied, yielding $b' : \neg D \in \B$. By the induction hypothesis, $rep(b') \in (\neg D)^{\I_\B}$. But $rep(b') = b' = rep(a')$, so the claim follows.
\item[(ii)] $a'$ is an $(r, E)$-proxy successor of $a$ for some concept $E$. By definition, $\{\neg D\mid a : \neg \exists r . D \in \B\} \subseteq \theory{\B}{a'}$, hence $a' : \neg D \in \B$. By the induction hypothesis, $rep(a') \in (\neg D)^{\I_\B}$, as required.
\end{itemize}\smallskip

\noindent\textbf{Case} $C = \iota D . E$\quad Assume $a : \iota D . E \in \B$. Since $\B$ is saturated, rule $(\iota_1^g)$ must have been applied, either introducing a fresh individual $a'$ such that $a' : D, a' : E \in \B$, or propagating $E$ to the least individual $a'$ already on $\B$ with $a' : D \in \B$. By the induction hypothesis, $rep(a') \in D^{\I_\B} \cap E^{\I_\B}$.

Suppose, for contradiction, that $rep(a) \notin (\iota D . E)^{\I_\B}$. Then there must exist some $a''$ such that $rep(a'') \neq rep(a')$ and $rep(a'') \in D^{\I_\B}$. Since $\B$ is saturated, rule $(cut_\iota^g)$ must have been applied to $\{a : \iota D . E, a'' : F\}$ for some concept $F$, resulting in either $a'' : D \in \B$ or $a'' : \neg D \in \B$.

In the first case, $a' : D, a'' : D \in \B$ implies $rep(a') = rep(a'')$, contradicting their assumed distinctness. In the second case, by the induction hypothesis, $rep(a'') \in (\neg D)^{\I_\B}$, so $rep(a'') \notin D^{\I_\B}$—again a contradiction.

\smallskip
\noindent\textbf{Case} $C = \neg \iota D . E$\quad Assume $a : \neg \iota D . E \in \B$. Suppose, for contradiction, that $rep(a) \notin (\neg \iota D . E)^{\I_\B}$. Then $rep(a) \in (\iota D . E)^{\I_\B}$, so there exists $rep(a') \in \Delta^{\I_\B}$ such that $\{rep(a')\} = D^{\I_\B}$ and $ rep(a')\in E^{\I_\B}$.

Since $\B$ is saturated, one of the following must hold:
\begin{itemize}
    \item[(i)] Rule $(\neg \iota^g)$ was applied to $a : \neg \iota D . E, a' : F$ for some concept $F$. Then one of the following holds:
    \begin{itemize}
        \item[(a)] $a' : \neg D \in \B$. By the induction hypothesis, $rep(a') \in (\neg D)^{\I_\B}$, hence $rep(a') \notin D^{\I_\B}$.
        \item[(b)] $a' : \neg E \in \B$. By the induction hypothesis, $rep(a') \in (\neg E)^{\I_\B}$, hence $rep(a') \notin E^{\I_\B}$.
        \item[(c)] Fresh individuals $b, b'$ were introduced such that $b : D, b : A_D^g$ and $b' : D, b' : \neg A_D^g \in \B$. Then, by the definition of $\cdot^{\I_\B}$ and the induction hypothesis, $rep(b) \in D^{\I_\B} \cap (A_D^g)^{\I_\B}$ and $rep(b') \in D^{\I_\B} \cap (\neg A_D^g)^{\I_\B}$, which implies $rep(b) \neq rep(b')$.
    \end{itemize}
    In all subcases, either $rep(a') \notin D^{\I_\B} \cap E^{\I_\B}$ or $D^{\I_\B}$ is not a singleton—contradicting the assumption that $\{rep(a')\}=D^{\I_\B}$ and $rep(a')\in E^{\I_\B}$.

    \item[(ii)] The application of $(\neg \iota^g)$ to $a : \neg \iota D . E$ was blocked. Then, by saturation, earlier applications introduced fresh individuals $b, b'$ with $b : D, b : A_D^g$ and $b' : D, b' : \neg A_D^g \in \B$. By the induction hypothesis and the definition of $\cdot^{\I_\B}$, $rep(b) \in D^{\I_\B} \cap (A_D^g)^{\I_\B}$ and $rep(b') \in D^{\I_\B} \cap (\neg A_D^g)^{\I_\B}$, so $rep(b) \neq rep(b')$, again contradicting the uniqueness of the $D$-satisfying individual.
\end{itemize}

\smallskip
\noindent\textbf{Case} $C = \{\iota D\}$\quad Assume $a : \{\iota D\} \in \B$. Since $\B$ is saturated, rule $(\iota_1^\ell)$ must have been applied, yielding $a : D \in \B$. By the induction hypothesis, $rep(a) \in D^{\I_\B}$.

Suppose, toward a contradiction, that $rep(a) \notin (\{\iota D\})^{\I_\B}$. Then there exists some $a''$ such that $rep(a'') \neq rep(a')$ and $rep(a'') \in D^{\I_\B}$. Since $\B$ is saturated, rule $(cut_\iota^\ell)$ must have been applied to $\{a : \{\iota D\}, a'' : F\}$ for some concept $F$, yielding either $a'' : D \in \B$ or $a'' : \neg D \in \B$.

In the first case, $a' : D, a'' : D \in \B$ implies $rep(a') = rep(a'')$, contradicting their distinctness. In the second, the induction hypothesis yields $rep(a'') \in (\neg D)^{\I_\B}$, hence $rep(a'') \notin D^{\I_\B}$—a contradiction.

\smallskip
\noindent\textbf{Case} $C = \neg \{\iota D\}$\quad Assume $a : \neg \{\iota D\} \in \B$. Since $\B$ is saturated, rule $(\neg \iota^\ell)$ must have been applied, yielding either $a : \neg D \in \B$, or $a : \neg A_D$, $b : D$, and $b : A_D \in \B$ for some individual $b$.

In the first case, the induction hypothesis gives $rep(a) \in (\neg D)^{\I_\B}$, so $rep(a) \notin D^{\I_\B}$. In the second, we have $rep(a) \in (\neg A_D)^{\I_\B}$ and $rep(b) \in D^{\I_\B} \cap (A_D)^{\I_\B}$, implying $rep(a) \neq rep(b)$. In both cases, $D^{\I_\B} \neq \{rep(a)\}$, so $rep(a) \notin (\{\iota D\})^{\I_\B}$, and thus $rep(a) \in (\neg \{\iota D\})^{\I_\B}$, as required.
\end{proof}

\noindent\textbf{Lemma~\ref{lem::NumberofPrefixesInBranch}.}\quad
Let $\T$ be a tableau for input $C$ (and $\mathcal{O}$), and let $\B$ be a branch of $\T$. Then the number of distinct individuals on $\B$ is bounded by $2^{4\cdot|C|}$ (or $2^{4\cdot(|C|+|\mathcal{O}|)} + k$, where $k$ is the number of individuals occurring in the ABox).

\begin{proof}
New individuals may be introduced by the rules $(ABox_\mathsf{I})$, $(ABox_r)$, $(\exists r)$, $(\iota_1^g)$, $(\neg\iota^g)$, and $(\neg\iota^\ell)$. Every non-root individual must be introduced by one of these rules. Define a function $f$ that maps each individual $a$ not introduced by any of the ABox rules to a set of concepts:
\begin{itemize}
    \item $\{D\}$, if $a$ is the root individual;
    \item $\{D, E\}$, if $a$ is introduced by $(\iota_1^g)$ on $a':\iota D.E$;
    \item $\{D, \pm A_D^g\}$, if introduced by $(\neg\iota^g)$ on $a':\neg\iota D.E$ and $a:\pm A_D^g\in\B$;
    \item $\{D, A_D\}$, if introduced by $(\neg\iota^\ell)$ on $a':\neg\{\iota D\}$;
    \item $\{D\} \cup \{\neg E \mid a':\neg\exists r.E \in \B\}$, if introduced by $(\exists r)$ on $a':\exists r.D$,
\end{itemize}
where $\pm A_E^g$ denotes either $A_E^g$ or $\neg A_E^g$. By construction, $f(a) \subseteq \theory{\B}{a}$. Thus, the range of $f$ is a subset of the power set of the set of concepts on $\B$, bounded by $2^{4\cdot|C|}$ (or $2^{4\cdot(|C| + |\mathcal{O}|)}$). Below we show that $f$ is injective, whence the bound on the number of non-ABox individuals follows. Adding the $k$ individuals from the ABox yields the claimed result.\smallskip
\noindent\textit{$f$ is injective.} Towards a contradiction, assume the opposite: there exist distinct non-ABox individuals $a, a'$ on $\B$ such that $f(a) = f(a')$. Without loss of generality, assume $a$ appears on $\B$ no later than $a'$. Then $a'$ must have been introduced by one of the following rules: $(\exists r)$, $(\iota_1^g)$, $(\neg\iota^g)$, or $(\neg\iota^\ell)$. We consider each case:\smallskip

\noindent\textbf{Case} $(\exists r)$\quad By definition, $f(a') = \{D\} \cup \{\neg E \mid a'':\neg\exists r.E \in \B \text{ at the time of application}\}$, where $a'':\exists r.D$ is the premise. Since $f(a) = f(a')$ and $a$ occurred earlier, for every concept $F\in f(a')$, $a:F$ was already present on $\B$ when $(\exists r)$ was applied. Hence, the blocking condition $(block_\exists)$ was satisfied, and the rule should not have been applied---a contradiction.\smallskip

\noindent\textbf{Case $(\iota_1^g)$}\quad The rule is applied to $a'':\iota D.E$, introducing $a'$. Then $f(a') = \{D, E\}$, and since $f(a) = f(a')$, we must have had $a:D$ and $a:E$ on $\B$ at the time. Thus, the introduction of new individual by $(\iota_1^g)$ was blocked---a contradiction.\smallskip

\noindent\textbf{Case $(\neg\iota^g)$}\quad Either $a'$ was introduced after $a$, or at the same time. In the latter case, $f(a) = f(a')$ would imply $\{D, A_D^g\} = \{D, \neg A_D^g\}$ for some $D$, which is impossible. In the former case, the rule is applied to $a'':\neg\iota D.E$, and $f(a') = \{D, A_D^g\}$ or $\{D, \neg A_D^g\}$. Assume the first (the second is analogous). Then $a:A_D^g$ must have been on $\B$ at the time, blocking the rule's application---a contradiction.\smallskip

\noindent\textbf{Case $(\neg\iota^\ell)$}\quad The rule was applied to $a'':\neg\{\iota D\}$, yielding $f(a') = \{D, A_D\}$. Since $f(a) = f(a')$ and $a$ occurred earlier, $a:A_D$ was already on $\B$, so the rule would only have introduced $a'':\neg A_D$ without adding a new prefix. Thus, $a'$ could not have been introduced---a contradiction.\smallskip

In all cases, assuming $f(a) = f(a')$ leads to a contradiction. Hence, $f$ is injective, as required.
\end{proof}

\section*{Additional technical details}

\noindent\textbf{Example C.}\quad
    Below we show that the $(cut_\iota^g)$ rule is essential for the completeness of $\TALCi$. Consider the following concept set:
\[
\{\iota \neg(A \sqcap A').\neg(A \sqcap A'),\quad \iota \neg A.A',\quad \iota \neg A'.A\}.
\]
This set is clearly unsatisfiable: the first concept enforces that exactly one individual satisfies $\neg(A \sqcap A')$, while the second and third introduce two distinct individuals satisfying $\neg A \sqcap A'$ and $\neg A' \sqcap A$, respectively---both satisfying $\neg(A \sqcap A')$, thus violating uniqueness.

Assuming $\TALCi$ is complete without $(cut_\iota^g)$, a tableau with the initial concept
\[
a: (\iota \neg(A \sqcap A').\neg(A \sqcap A')) \sqcap (\iota \neg A.A') \sqcap (\iota \neg A'.A)
\]
should close. However, without $(cut_\iota^g)$, the tableau construction yields the following (initial applications of $(\sqcap)$ are omitted):

    \begin{center}
    \begin{scriptsize}
    \begin{tikzpicture}
     \node (a) at (0,0) {$a:\iota \neg(A\sqcap A').\neg(A\sqcap A'),\quad a:\iota\neg A.A', \quad a:\iota\neg A'.A$};

     \node (b) at (0,-.7) {$b:\neg(A\sqcap A')$};

     \node (c) at (0,-1.4) {$c:\neg A$,\quad $c:A'$};

     \node (d) at (0,-2.1) {$d:\neg A'$,\quad $d:A$};

     \node (e) at (-2.5,-2.8) {$b:\neg A$};

     \node (f) at (-2.5,-3.5) {$b:A'$,\quad $c:\neg(A\sqcap A')$};

     \node (f1) at (-3.5,-4.2) {$c:\neg A$};

     \node (f2) at (-1.5,-4.2) {$c:\neg A'$};

     \node (g) at (2.5,-2.8) {$b:\neg A'$};

     \node (h) at (2.5,-3.5) {$b:A$,\quad $d:\neg(A\sqcap A')$};

    \node (h1) at (1.5,-4.2) {$d:\neg A$};

     \node (h2) at (3.5,-4.2) {$d:\neg A'$};

     \node (f1o) at (-3.5,-4.9) {open};

    \node (f2c) at (-1.5,-4.9) {$\bot$};

    \node (h2o) at (3.5,-4.9) {open};

    \node (h1c) at (1.5,-4.9) {$\bot$};

    \node (aca) at (-4.5,-5.6) {$d:\neg(A\sqcap A')$};

    \node (acb) at (-2.5,-5.6) {$d:\neg\neg(A\sqcap A')$};

    \node (acc) at (-4.5,-6.3) {$c:A$};

    \node (acd) at (-2.5,-6.3) {$d:A'$};

    \node (ace) at (-4.5,-7) {$\bot$};

    \node (acf) at (-2.5,-7) {$\bot$};

    \node (aca') at (4.5,-5.6) {$c:\neg\neg(A\sqcap A')$};

    \node (acb') at (2.5,-5.6) {$c:\neg(A\sqcap A')$};

    \node (acc') at (4.5,-6.3) {$c:A$};

    \node (acd') at (2.5,-6.3) {$c:A$};

    \node (ace') at (4.5,-7) {$\bot$};

    \node (acf') at (2.5,-7) {$\bot$};

    \draw[thick] (a) to node[right] {\tiny$(\iota_1^g)$} (b) to node[right] {\tiny$(\iota_1^g)$} (c) to node[right] {\tiny$(\iota_1^g)$} (d) to node[right,xshift=28pt] {\tiny$(\neg\sqcap)$} (e) to node[right] {\tiny$(\iota_1^g)$} (f) to node[right,yshift=-3pt,xshift=3pt] {\tiny$(\neg\sqcap$)} (f1) -- (f1o);
    \draw[thick] (f) -- (f2) to node[left] {\tiny$(\bot)$} (f2c);
    \draw[thick] (d) -- (g) to node[left] {\tiny$(@_2)$} (h) to node[right,yshift=-3pt,xshift=3pt] {\tiny$(\neg\sqcap)$} (h1) to node[right] {\tiny$(\bot)$} (h1c);
    \draw[thick] (h) -- (h2) -- (h2o);

    \draw[thick] (f1o) to node[right,yshift=-3pt] {\tiny$(cut_\iota^g)$} (aca) to node[left] {\tiny$(\iota_2^g)$} (acc) to node[left] {\tiny$(\bot)$} (ace);

    \draw[thick] (f1o) -- (acb) to node[right] {\tiny$(\neg\neg)$, $(\sqcap)$} (acd) to node[right] {\tiny$(\bot)$} (acf);

    \draw[thick] (h2o) to node[left,yshift=-3pt] {\tiny$(cut_\iota^g)$} (aca') to node[right] {\tiny$(\neg\neg)$, $(\sqcap)$} (acc') to node[right] {\tiny$(\bot)$} (ace');

    \draw[thick] (h2o) -- (acb') to node[left] {\tiny$(\iota_2^g)$} (acd') to node[left] {\tiny$(\bot)$} (acf');

    \draw[dashed,very thick] (-5.5,-5.05) to node[above,midway] {without $(cut_\iota^g)$} node[below,midway] {with $(cut_\iota^g)$} (5.5,-5.05);
    \end{tikzpicture}
    \end{scriptsize}
    \end{center}
    As shown, without $(cut_\iota^g)$, some branches remain open. Only by applying $(cut_\iota^g)$ can we close them, demonstrating the rule's necessity for completeness.\medskip

\noindent\textbf{Proposition~D.}\quad
There exists a satisfiable $\ALCi$ concept $C$ such that every model of $C$ contains a path of exponential length in $|C|$.

\begin{proof}
We prove the proposition by encoding a \emph{binary counter} in $C$, a technique previously employed by, e.g., Blackburn et al.~\cite{blackburn2002modal}.

Let $A_1, \ldots, A_n$ be atomic concepts representing bits in an $n$-bit binary number, with $A_1$ as the least significant and $A_n$ as the most significant bit. Each $A_i$ denotes that the $i$th bit is $1$, while $\neg A_i$ denotes that it is $0$.

To increment a binary number, we flip the least significant $0$ to $1$ and reset all preceding $1$'s to $0$, leaving other bits unchanged. For example, if the least significant bit is $0$, it suffices to flip that bit alone:
\[
D_1 := \neg A_1 \sqcup \left[\forall r. A_1 \sqcap \bigsqcap_{i=2}^n \left((\neg A_i \sqcup \forall r. A_i) \sqcap (A_i \sqcup \forall r. \neg A_i)\right)\right].
\]
More generally, for $i > 1$, the increment step is:
\[
D_i := \neg A_i \sqcap \bigsqcap_{j=1}^{i-1} \neg A_j \sqcup \left[\forall r.\left(A_i \sqcap \bigsqcap_{j=1}^{i-1} \neg A_j\right) \sqcap \bigsqcap_{j=i+1}^n \left((\neg A_j \sqcup \forall r. A_j) \sqcap (A_j \sqcup \forall r. \neg A_j)\right)\right].
\]

We define:
\[
D := \bigsqcap_{i=1}^n D_i \sqcap \exists r.\top.
\]
This concept enforces the correct binary increment behavior and guarantees the existence of an $r$-successor where the incremented number holds.

We now define the concept $C$:
\[
C := \iota \neg D. \bigsqcap_{i=1}^n A_i \sqcap \iota \bigsqcap_{i=1}^n \neg A_i. \top.
\]
The first conjunct ensures that there is a unique individual $a^*$ not satisfying $D$, and that $a^*$ encodes $2^n - 1$ in binary. The second conjunct ensures there is a unique individual $a_0$ encoding $0$. Since these encodings differ, we have $a_0 \neq a^*$.

As $D$ holds at $a_0$, the increment process begins there, generating $2^n - 3$ successors (as $a_0$ and $a$ already exist), each representing the next binary number. Since all these individuals (except $a^*$) satisfy $D$, the rule is repeatedly applied. This enforces the existence of an $r$-path of length $2^n$ in any model of $C$:

\begin{center}
\begin{tikzpicture}
    \node[circle,draw=black,fill=black,minimum size=3pt,inner sep=0,outer sep=2pt,label={90:\scriptsize$\neg A_1,\neg A_2,\ldots,\neg A_n$}] (a) at (0,0) {};
    \node[circle,draw=black,fill=black,minimum size=3pt,inner sep=0,outer sep=2pt,label={90:\scriptsize$A_1,\neg A_2,\ldots,\neg A_n$}] (b) at (3,0) {};
    \node[draw=none,fill=none,outer sep=2pt] (c) at (6,0) {$\cdots$};
    \node[circle,draw=black,fill=black,minimum size=3pt,inner sep=0,outer sep=2pt,label={90:\scriptsize$A_1,A_2,\ldots,A_n$}] (d) at (9,0) {};
    \draw[very thick,->] (a) to[bend right] node[above] {$r$} (b);
    \draw[very thick,->] (b) to[bend right] node[above] {$r$} (c);
    \draw[very thick,->] (c) to[bend right] node[above] {$r$} (d);
\end{tikzpicture}
\end{center}

This structure must be embedded in every model of $C$. Since the size of $C$ is in $\mathcal{O}(n^2)$, the guaranteed path length is exponential in $|C|$, as required.
\end{proof}

\end{document}